\documentclass[noshowkeys,floatfix,twocolumn,nofootinbib,aps,prb,10pt, superscriptaddress]{revtex4-2}
\usepackage{filecontents}
\usepackage[utf8]{inputenc} 
\usepackage[german,english]{babel} 
\usepackage[normalem]{ulem} 
\usepackage{amsfonts} 
\usepackage{amsmath} 
\usepackage{amssymb} 
\usepackage{bbm} 
\usepackage{physics} 
\usepackage{braket} 
\usepackage{bm} 
\usepackage{cases} 
\usepackage{tikz}
\usetikzlibrary{positioning,decorations.pathmorphing, decorations.pathreplacing, decorations.shapes}
\usepackage[compat=1.1.0]{tikz-feynhand}
\usepackage{calc}
\usepackage{graphicx} 
\usepackage{xcolor} 
\usepackage[caption=false,position=top]{subfig}
\usepackage{xr}
\usepackage[citecolor=blue, linkcolor=blue, urlcolor=blue, filecolor=blue, hypertexnames=true, colorlinks, bookmarksopenlevel=1, pdfstartview=FitH, bookmarksopen, bookmarksnumbered, plainpages=false, linktocpage]{hyperref}
\let\oldciteauthor=\citeauthor
\def\citeauthor#1{{\hypersetup{citecolor=black}\oldciteauthor{#1}}}
\let\oldfootnote=\footnote
\def\footnote#1{\hypersetup{citecolor=blue}\oldfootnote{#1}}
\usepackage[capitalize]{cleveref}
\usepackage{url}

\def\cal#1{\mathcal{#1}}

\newcommand{\opc}{\hat{c}}

\newcommand{\T}[1]{\textrm{#1}}

\newcommand{\up}{\uparrow}

\newcommand{\bk}{\bm{k}}
\newcommand{\bq}{\bm{q}}
\newcommand{\bpos}{\bm{r}}
\newcommand{\bG}{\bm{G}}

\newcommand{\Pm}{\mathcal{P}}
\newcommand{\Qm}{\mathcal{Q}}

\newcommand{\Fm}{\mathcal{F}}

\newcommand{\nl}{\nonumber\\}

\tikzset{
    between/.style args={#1 and #2}{
         at = ($(#1)!0.5!(#2)$)
    }
}
\makeatletter
\pgfdeclaredecoration{bettercoil}{coil}
{
  \state{coil}[switch if less than=%
    0.5\pgfdecorationsegmentlength+
    \pgfdecorationsegmentaspect\pgfdecorationsegmentamplitude+%
    \pgfdecorationsegmentaspect\pgfdecorationsegmentamplitude to last,
               width=+\pgfdecorationsegmentlength]
  {
    \pgfpathcurveto
    {\pgfpoint@oncoil{0    }{ 0.555}{1}}
    {\pgfpoint@oncoil{0.445}{ 1    }{2}}
    {\pgfpoint@oncoil{1    }{ 1    }{3}}
    \pgfpathcurveto
    {\pgfpoint@oncoil{1.555}{ 1    }{4}}
    {\pgfpoint@oncoil{2    }{ 0.555}{5}}
    {\pgfpoint@oncoil{2    }{ 0    }{6}}
    \pgfpathcurveto
    {\pgfpoint@oncoil{2    }{-0.555}{7}}
    {\pgfpoint@oncoil{1.555}{-1    }{8}}
    {\pgfpoint@oncoil{1    }{-1    }{9}}
    \pgfpathcurveto
    {\pgfpoint@oncoil{0.445}{-1    }{10}}
    {\pgfpoint@oncoil{0    }{-0.555}{11}}
    {\pgfpoint@oncoil{0    }{ 0    }{12}}
  }
  \state{last}[next state=final]
  {
    \pgfpathcurveto
    {\pgfpoint@oncoil{0    }{ 0.555}{1}}
    {\pgfpoint@oncoil{0.445}{ 1    }{2}}
    {\pgfpoint@oncoil{1    }{ 1    }{3}}
    \pgfpathcurveto
    {\pgfpoint@oncoil{1.555}{ 1    }{4}}
    {\pgfpoint@oncoil{2    }{ 0.555}{5}}
    {\pgfpoint@oncoil{2    }{ 0    }{6}}
  }
  \state{final}{}
}

\def\pgfpoint@oncoil#1#2#3{%
  \pgf@x=#1\pgfdecorationsegmentamplitude%
  \pgf@x=\pgfdecorationsegmentaspect\pgf@x%
  \pgf@y=#2\pgfdecorationsegmentamplitude%
  \pgf@xa=0.083333333333\pgfdecorationsegmentlength%
  \advance\pgf@x by#3\pgf@xa%
}
\makeatother

\begin{document}


\date{\today}
\title{Friedel oscillations and chiral superconductivity in monolayer NbSe\texorpdfstring{$_2$}{2}}
\author{Julian Siegl}
\email{Julian.Siegl@ur.de}
\author{Anton Bleibaum}
\affiliation{Institute for Theoretical Physics, University of Regensburg, 93 053 Regensburg, Germany}
\author{Wen Wan}
\affiliation{Donostia International Physics Center, Paseo Manuel de Lardizábal 4, 20018 San Sebastián, Spain.}
\author{Marcin Kurpas}
\affiliation{Institute of Physics, University of Silesia in Katowice, 41-500 Chorzów, Poland}
\author{\hbox{John Schliemann}}
\affiliation{Institute for Theoretical Physics, University of Regensburg, 93 053 Regensburg, Germany} 
\author{Miguel M. Ugeda}
\affiliation{Donostia International Physics Center, Paseo Manuel de Lardizábal 4, 20018 San Sebastián, Spain.}
\affiliation{Centro de Física de Materiales, Paseo Manuel de Lardizábal 5, 20018 San Sebastián, Spain.}
\affiliation{Ikerbasque, Basque Foundation for Science, Bilbao 48013, Spain}
\author{Magdalena Marganska}
\affiliation{Institute for Theoretical Physics, University of Regensburg, 93 053 Regensburg, Germany}
\affiliation{Department of Theoretical of Physics, Wrocław University of Science and Technology, Wybrzeże Wyspiańskiego 27, 50-370 Wrocław, Poland}
\author{Milena Grifoni}
\email{Milena.Grifoni@ur.de}
\affiliation{Institute for Theoretical Physics, University of Regensburg, 93 053 Regensburg, Germany}
\begin{abstract}
In 1965 Kohn and Luttinger proposed a genuine electronic mechanism for superconductivity.
Despite the bare electrostatic interaction between two electrons being repulsive, in a metal electron-hole fluctuations can give rise to Friedel oscillations of the screened Coulomb potential.
Cooper pairing among the electrons then emerges when taking advantage of the attractive regions.
The nature of the leading pairing mechanism in some two-dimensional transition metal dichalcogenides is still debated.
Focusing on NbSe$_2$, we show that superconductivity can be induced by the Coulomb interaction when accounting for screening effects on the trigonal lattice with multiple orbitals.
Using ab initio-based tight-binding parametrizations for the relevant low-energy $d$-bands, we evaluate the screened interaction microscopically, in a scheme  including  Bloch overlaps.
In the direct space, we find long-range Friedel oscillations alternating in sign, a key to the Kohn-Luttinger mechanism.
The momentum-resolved gap equations predict two degenerate  solutions at the critical temperature $T_{\T{c}}$, signaling the unconventional nature of the pairing.
Their complex linear combination, i.e., a chiral gap with $p$-like symmetry, provides the  ground state of the system. 
 Our prediction of a fully gapped chiral phase well below $T_{\T{c}}$ is  in excellent agreement with the spectral function extracted from tunneling spectroscopy measurements of single-layer NbSe$_2$. 
\end{abstract}
\maketitle

In conventional superconductors, electron pairing arises from virtual phonon exchange~\cite{bardeenTheorySuperconductivity1957}.
In this case, the slowly moving ions mediate an effective attractive interaction that allows the electrons to overcome their mutual repulsion and form Cooper pairs.
This mechanism favors $s$-wave pairing, where the total spin of the electron-pair is zero, and the gap function is isotropic in momentum space.
Kohn and Luttinger have proposed an alternative pairing mechanism originating from pure Coulomb repulsion~\cite{kohnNewMechanismSuperconductivity1965}.
It is mediated by electron-hole fluctuations of the metal, which result in long-range Friedel oscillations~\cite{friedelXIVDistributionElectrons1952} of the screened Coulomb interaction. For conventional superconductors, however, the effect is subleading compared to phonon-induced pairing~\cite{kohnNewMechanismSuperconductivity1965}. As we shall show, it can yield gap sizes compatible with the experiment in the compound investigated in this work, monolayer NbSe$_2$. 

When superconductivity emerges from screened repulsion, the free energy is minimized by an anisotropic gap function~\cite{scalapinoCommonThreadPairing2012}, and the pairing symmetry is associated to irreducible representations of the crystal.
If the symmetry of the Bravais lattice allows a two-dimensional representation, Coulomb repulsion can even mediate a superconducting chiral phase, with spontaneous time-reversal symmetry breaking~\cite{kallinChiralSuperconductors2016}.
Two-dimensional (2D) systems with trigonal lattice are potential candidates for becoming chiral superconductors~\cite{mackenzieEvenOdderTwentythree2017,black-schafferChiralDwaveSuperconductivity2014,nandkishoreChiralSuperconductivityRepulsive2012}.
Evidence for chiral $d$-wave superconductivity has recently been reported for tin monolayers on a Si(111) surface~\cite{mingEvidenceChiralSuperconductivity2023}, and a transition from a nematic to a chiral phase has been proposed for TaS$_2$~\cite{ribakChiralSuperconductivityAlternate2020,silberTwocomponentNematicSuperconductivity2024,wanUnconventionalSuperconductivityChiral2024}. 
Here we microscopically
investigate the nature of superconductivity in
monolayer NbSe$_2$,  
a 2D compound of the family of transition metal dichalcogenides (TMDs) with Nb atoms arranged on a trigonal lattice.
It combines strong Ising spin-orbit coupling (SOC)~\cite{xuSpinPseudospinsLayered2014,xiIsingPairingSuperconducting2016,zhangIsingPairingAtomically2021} with superconductivity~\cite{caoQualityHeterostructuresTwoDimensional2015,ugedaCharacterizationCollectiveGround2016,wanObservationSuperconductingCollective2022}, leading to a violation of the Pauli limit in strong in-plane magnetic fields~\cite{frigeriSuperconductivityInversionSymmetry2004,xiGateTuningElectronic2016,ilicEnhancementUpperCritical2017,mockliMagneticfieldInduced2019,shafferCrystallineNodalTopological2020,choNodalNematicSuperconducting2022}, and an incommensurate charge density wave~\cite{johannesFermisurfaceNestingOrigin2006,xiStronglyEnhancedChargedensitywave2015,xiGateTuningElectronic2016}.
While bulk NbSe$_2$ is commonly assumed to be a conventional phonon-mediated superconductor~\cite{heilOriginSuperconductivityLatent2017,sannaRealspaceAnisotropySuperconducting2022}, the mechanism for the monolayer limit is still actively discussed and ranges from phonon-mediated \cite{zhengElectronphononCouplingCoexistence2019,dasElectronphononCouplingSpin2023} to more exotic pairings~\cite{wickramaratneIsingSuperconductivityMagnetism2020,shafferCrystallineNodalTopological2020,horholdTwobandsIsingSuperconductivity2023,royUnconventionalPairingIsing2024}.
Our findings, based on a microscopic calculation of the screened interaction, support the unconventional nature of the superconductivity in the monolayer case.

The expected enhanced importance of unconventional pairing in 2D TMDs arises from their fragmented Fermi surface, cf.~\cref{fig:bands}, which implies distinct intervalley and intravalley scattering processes, and whose amplitude can be large due to the weaker screening in lower dimensions~\cite{hybertsenElectronCorrelationSemiconductors1986,rohlfingElectronholeExcitationsOptical2000}.
Indeed, model calculations have shown that an effective attraction from competing repulsive interactions requires dominance of the short-range intervalley scattering over the long-range intravalley one~\cite{roldanInteractionsSuperconductivityHeavily2013,shafferCrystallineNodalTopological2020,horholdTwobandsIsingSuperconductivity2023}. 
In this work, we show that a generalized random phase approximation (RPA) on the trigonal lattice of Nb atoms yields a selectively screened Coulomb interaction, that favors short-range intervalley scattering of Cooper pairs.
The multi-orbital character of the valence band, the Umklapp processes as well as the spatial distribution of the atomic $d$-orbitals play an important role in this context.
In the real space the screened potential exhibits Friedel oscillations with alternating sign, thus 
providing a realization of the pairing mechanism based on electron-hole fluctuations proposed by Kohn and Luttinger~\cite{kohnNewMechanismSuperconductivity1965}.
Upon solving the coupled linearized gap equations for the full screened potential, we find  two degenerate solutions near $T_{\T{c}}$. By solving self-consistency equations, we predict the gapped $p$-wave chiral  solution to provide the stable phase down to the lowest temperatures. 
Our findings are in very good agreement with low-temperature tunneling spectroscopy measurements acquired on monolayer NbSe$_2$. 
\subsection*{Band structure and orbital composition}
\begin{figure}[t]
\includegraphics[width=\columnwidth]{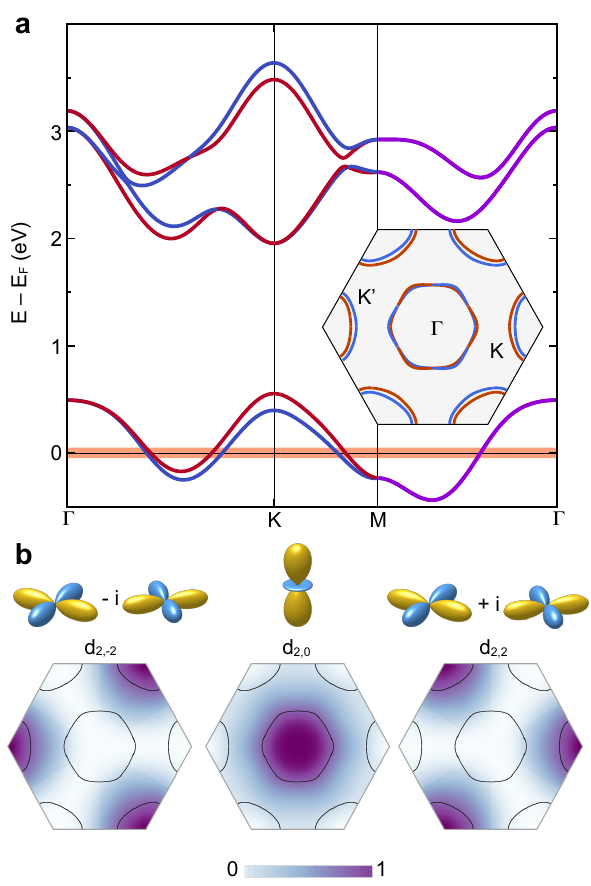}
 \caption{
    {\bf{Band structure, Fermi surfaces and orbital contributions in monolayer
    NbSe$_{2}$. a}}
    Tight-binding Bloch bands and Fermi surfaces (inset).
    The Fermi energy lies within the spin split valence bands; the spin degeneracy is removed by the Ising spin-orbit coupling, resulting in spin-resolved Fermi surfaces.
    {\bf{b}} Sketch of the three atomic $d_{l,m}$ orbitals used in the calculation and their contribution to the composition of the valence bands.
    The quantum numbers $l,m$ denote the total angular momentum of the orbital and its \hbox{azimuthal} projection.  
    }
    \label{fig:bands}
\end{figure}

The starting point for our considerations is a tight-binding model for the Bloch band structure of monolayer NbSe\texorpdfstring{$_2$}{2} obtained by using the three most relevant  orbitals of Nb, $d_{z^2}=d_{2,0}$ and $d_{2,\pm 2} = d_{x^2-y^2} \pm i d_{xy}$~\cite{liuThreebandTightbindingModel2013}, cf. Fig. \ref{fig:bands}, fitted to ab-initio calculations.
Here the parametrization by \citeauthor{heMagneticFieldDriven2018}~\cite{heMagneticFieldDriven2018} is used, but similar results are obtained from the one reported by \citeauthor{kimQuasiparticleEnergyBands2017}~\cite{kimQuasiparticleEnergyBands2017}.
There are two metallic valence bands relevant for superconducting pairing at the Fermi level, with distinct spin and which are split by Ising SOC~\cite{liuThreebandTightbindingModel2013}.
As seen in the inset of Fig. \ref{fig:bands}a, multiple disjoint Fermi surfaces are present, which suggests multi-patch superconductivity similar to the one arising in the cuprates and iron pnictides~\cite{peraliTwogapModelUnderdoped2000,maitiSuperconductivityRepulsiveInteraction2013}.
While the planar $d$-orbitals mostly contribute to the band composition around the $K$ and $K$' valleys, $d_{2,0}$ is dominant at the $\Gamma$ valley, see Fig.~\ref{fig:bands}b. 
 \begin{figure*}[ht]
    \includegraphics[width=\textwidth]{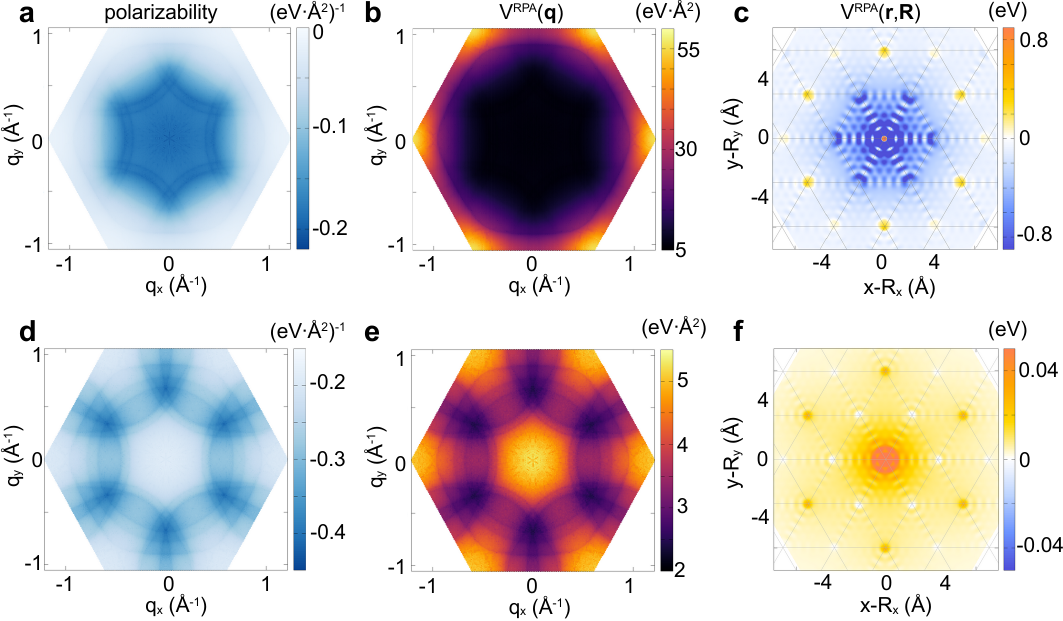}
    \caption{
        {\bf{Macroscopic polarizability and screened interaction in the reciprocal and direct space. a}}
        The macroscopic polarizability $P_{\bm{0},\bm{0}}$ displays a plateau for small momentum transfers $\bq \approx 0$ ($\bm{0}=:\bq_\Gamma$), where its magnitude $\vert P_{\bm{0},\bm{0}} \vert$ attains its maximal value, and decreases for large momentum transfers $\bq\approx\pm\boldsymbol{K}$ ($\boldsymbol{K}=:\bq_K$).
        This behavior reflects the strong momentum dependence of the Bloch overlaps.
        {\bf{b}} The effective screened interaction is correspondingly strongly suppressed for small momentum transfer and exhibits maxima around $\pm\bq_K$.
        {\bf{c}} In real space the screened potential displays Friedel oscillations which bear signatures of the underlying trigonal lattice.
        The potential is strongly repulsive on-site, but attractive at the nearest neighbor sites. 
        {\bf{d,e,f}} Same observables as in the top panels evaluated in constant matrix element approximation, i.e., setting the Bloch overlaps to 1.
        In this case, the screening at large momentum transfers is overestimated and hence the screened interaction is largest around $\bm{q}_{\Gamma}$.
        In turn, the oscillations in real space are weak and the potential remains repulsive everywhere except small regions of weak attraction. 
    } 
    \label{fig:screened}
\end{figure*}
\subsection*{Screened Coulomb interaction\label{sec: screened couloumb}}
In crystalline systems the screened interaction is in  general a function of both positions of the two interacting charges. It  can be expanded as~\cite{hybertsenElectronCorrelationSemiconductors1986}
\begin{align}
    &V(\bm{r},\bm{r}')
    &=\frac{1}{\cal{V}}\sum_{\bq',\bm{G},\bm{G}'}V^{}_{\bm{G},\bm{G}'}(\bq')e^{i(\bq'+\bm{G})\cdot\bpos}e^{-i(\bq'+\bm{G}')\cdot\bpos'}\,,
    \label{screened interaction real space}
\end{align}
where $\bG$, $\bm{G}'$ are reciprocal lattice vectors, $\bq'$ is a crystal momentum not confined to lie in-plane, and $\cal{V}$ is the sample volume.
An exact evaluation of the screened interaction is  not possible.
However, a closed form can be obtained in the random phase approximation (RPA), in which the polarizability is approximated by that of independent Bloch electrons~\cite{hybertsenElectronCorrelationSemiconductors1986}.
With a focus on NbSe$_2$, we view the orbitals as tightly confined to the plane of the material, cf. Fig.~\ref{fig:bands}. This enables us to obtain a two-dimensional version of the RPA-screened interaction which properly accounts for the long-range behavior.
The resulting two-dimensional screened interaction $V_{\bm{G},\bm{G}'}^{\T{2D,RPA}}(\bq)$,  cf. the Supplementary Material and \cref{sm-two dimensional screened interaction tensor} therein, can be expressed in terms of the 2D unscreened one, $V_{0}^{\T{2D}}(\bq)=\frac{e^2}{2\epsilon_0 q}$, as 
\begin{align}
    \label{formal solution dyson equation in 2d}
    &V_{\bm{G},\bm{G}'}^{\T{2D,RPA}}(\bq;\omega)=\epsilon^{-1}_{\bG,\bG'}(\bq;\omega)V_{0}^{\T{2D}}(\bq+\bG')\,,
\end{align}
with $\epsilon^{-1}_{\bG,\bG'}$ the inverse of the dielectric tensor
\begin{align}
   &\epsilon_{\bG,\bG'}(\bq;\omega)=\delta_{\bG,\bG'}-V^{\T{2D}}_0(\bq+\bG)P_{\bG,\bG'}(\bq;\omega)\,.
   \label{dielectric tensor}
\end{align}
The elements of the polarizability tensor in \cref{dielectric tensor} are given by
\begin{align}
   \!\! P_{\bG,\bG'}(\bm{q};\omega)=\sum_{\bm{p},\sigma}\chi_\sigma(\bm{p}+\bq,\bm{p};\omega)
    \Fm^{\sigma}_{\bm{p},\bm{p}+\bq}(\bG)\Fm^{\sigma}_{\bm{p}+\bq,\bm{p}}(-\bG')\,,
    \label{susceptibility}
\end{align}
with $\chi_\sigma$ accounting for electron-hole fluctuations, cf.~\cref{susceptibility sigma}, and   the Bloch overlaps $\Fm$ defined as
\begin{align}
    \Fm^{\sigma}_{\bk,\bk'}(\bm{G}) =&\int_{\cal{V}_p}d\bm{r}e^{-i\bm{G}\cdot\bm{r}}\bm{u}^\dagger_{\sigma,\bk}(\bm{r})\bm{u}_{\sigma,\bk'}(\bm{r})\,.
    \label{overlap factors}
\end{align}
Here $\bm{u}_{\sigma,\bk}$ are Bloch spinors, i.e., the periodic part of the Bloch functions holding the information about their orbital composition, and ${\cal{V}_p}$ is the volume of the unit cell.
%
We show in Fig.~\ref{fig:screened}a the macroscopic static polarizability, the element $P_{\bm{0},\bm{0}}({\bf{q}},0)$ of the polarizability tensor, properly accounting for the Bloch overlaps.
In Fig.~\ref{fig:screened}d the same quantity is shown in the constant matrix element approximation~\cite{johannesFermisurfaceNestingOrigin2006,kimQuasiparticleEnergyBands2017}, where such overlaps are set to unity. 
In both cases a plateau is observed for small momentum transfers, $\bq\approx\bq_\Gamma$.
However, while the magnitude $\vert P_{\bm{0},\bm{0}}\vert $ of the full macroscopic polarizability is \textit{maximal} there, it is \textit{minimal} for the approximated one.
As such, retaining the directionality of the Bloch overlaps in momentum space, cf.~\cref{sm-fig: form factors general} in the Supplementary Material, is crucial to correctly account for the long-range behavior of the screened interaction.
This qualitative difference is visualized in Figs.~\ref{fig:screened}b and ~\ref{fig:screened}e where we display the effective screened interaction potential $V^{\T{RPA}}(\bm{q}) := \sum_{\bG,\bG'}V_{\bm{G},\bm{G}'}^{\T{2D,RPA}}(\bm{q})$  calculated with and without the constant matrix element approximation for the polarizability, respectively (see Methods). 
The effective RPA interaction potential shown in Fig.~\ref{fig:screened}b is strongly suppressed for $\bq\approx\bq_\Gamma$, and displays maxima at $\pm\bq_K$.
The strong suppression for low momentum exchanges $\bm{q}$ supports the scenario of dominant intervalley scattering discussed by \citeauthor{roldanInteractionsSuperconductivityHeavily2013,horholdTwobandsIsingSuperconductivity2023}~\cite{roldanInteractionsSuperconductivityHeavily2013,horholdTwobandsIsingSuperconductivity2023}.
In constant matrix element approximation, the effective interaction potential is in contrast largest at the $\bq_\Gamma$ plateau, see Fig.~\ref{fig:screened}e, while being in general smaller throughout the whole Brillouin zone.
Hence, the impact of screening for large momentum electron-hole fluctuations is overestimated in the constant matrix element approximation.
When turning to the real space, we find that $V^{\text {RPA}}(\bm{r},\bm{r}')$, calculated according to Eq.~(\ref{screened interaction real space}), displays sizable Friedel oscillations.
The interference pattern originates from the superposition of contributions from $\bq_K$ and $-\bq_K$ regions and is a signature of the trigonal lattice.
As seen in Fig.~\ref{fig:screened}c, the potential as a function of $\bm{r}$, with $\bm{r}'=\bm{R}$ fixed to an Nb site, exhibits an on-site repulsion at the origin and repulsive regions around the next nearest neighbor lattice sites, while it is strongly attractive at the nearest neighbor sites. 
This behavior is contrasted by the outcome of the constant matrix element approximation, where the screened potential remains repulsive almost everywhere with only narrow regions of weak attraction, see Fig.~\ref{fig:screened}f.
This finding constitutes a major result of the present work. 

As we demonstrate  below, in a way similar to the Friedel oscillations mechanism proposed by Kohn and Luttinger for simple metals~\cite{kohnNewMechanismSuperconductivity1965}, the interplay of attractive and repulsive oscillations of the screened potential can provide the leading mechanism for superconductivity  in monolayer NbSe$_{2}$.
\subsection*{Gap symmetries and temperature dependence~\label{sec: gap symmetries}}
We next discuss the properties of the superconducting instability mediated by the screened potential. 
In general, condensates may be formed with finite center-of-mass momentum for Cooper pairs.
However, the matrix elements of the screened interaction contributing to pairing are in general complex, with a strongly fluctuating phase resulting from the product of Bloch overlaps, cf.~\cref{matrix element bloch states expanded}.
This tends to favor pairing between time-reversed states, whose Bloch spinors obey the relation $\bm{u}_{\bar{\sigma},\bar{\bk}} = \bm{u}_{\sigma,{\bk}}^*$, with  $-\bm{k}=:\bar{\bm{k}},-\sigma=:\bar{\sigma}$,  resulting in a summation of real and positive contributions to the pairing. 
We hence retain only the scattering between time reversal partners ($\bm{k},\sigma;\bar{\bm{k}}, \bar{\sigma}$), and consider in the following the interaction matrix elements 
${V_{\bk,\bk',\sigma}:=\bra{\bk',\sigma;\bar{\bk}',\sigma'}\hat{V}\ket{\bk,\sigma;\bar{\bk},\bar{\sigma}}}$ involving Kramers pairs of Bloch states. 
This restriction to  time-reversal pairs yields a low energy Hamiltonian of the typical BCS form~\cite{bardeenTheorySuperconductivity1957}.
Performing a mean field approximation, in turn, leads to the familiar BCS gap equation 
\begin{align}
    \label{gap equation trs}
    &\Delta_{\bk,\sigma}=-\sum_{\bk'}V_{\bk,\bk',\sigma}\Pi(E_{\bk,\sigma})\Delta_{\bk',\sigma}\,,
\end{align}
where the pairing function $\Pi(E)=\tanh(\beta E/2)/2E$ depends on the gap through the quasiparticle energies
$E_{\bk,\sigma}=\sqrt{\xi_{\bk,\sigma}^2+\vert\Delta_{\bk,\sigma}\vert^2}$, and on the inverse temperature  $\beta=1/k_BT$. 
To access observables near the phase transition, we linearize~\cref{gap equation trs} by replacing the excitation energies $E_{\bk,\sigma}$ in the pairing function by the single-particle energies $\xi_{\bk,\sigma}$ and solving the resulting linear problem in a range of $\pm \Lambda$  around the Fermi energy~\cite{shafferCrystallineNodalTopological2020,royUnconventionalPairingIsing2024}, cf. Methods. 
\begin{figure}[t]
    \includegraphics[width=\columnwidth]{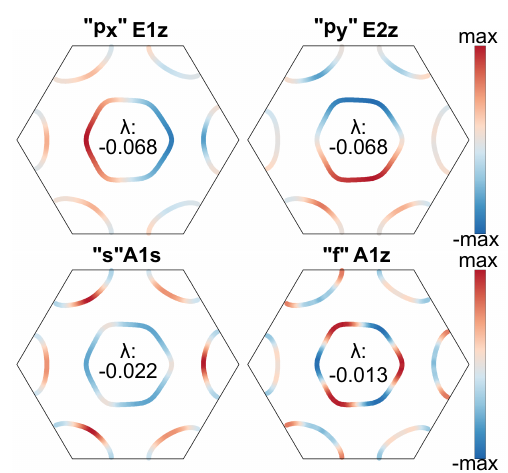}
    \caption{
    {\bf{Leading instabilities as found from solving the linearized gap equations.}}
    The solutions are labeled by a name (in quotation marks) followed by a label indicating which is the dominant irreducible representation contributing to it, as well as its majority component being either spin singlet (s) or triplet z-component (z).
    Lower eigenvalue $\lambda$ indicates larger T$_c$ of the corresponding instability. 
    The two solutions with highest critical temperature are degenerate,  provided mostly by the z-triplet components of the $E'$ irrep of the $D_{\mathrm{3h}}$ group.
    The two following solutions are combinations of the $A'_1$ $z$-triplet and $A'_1$ singlet with the predominantly singlet pairing having a higher critical temperature.
    }
    \label{fig: linearized gaps}
\end{figure}
The leading instabilities with the highest $T_{\T{c}}$ are shown in \cref{fig: linearized gaps} for the parameter set by \citeauthor{heMagneticFieldDriven2018}~\cite{heMagneticFieldDriven2018} and a truncation to third nearest neighbors of the 1BZ in the calculation of the matrix $V^{\T{2D,RPA}}_{\bG,\bG'}$.
They belong to irreducible representations of the symmetry group of the material, which for NbSe$_2$ is $D_{\mathrm{3h}}$.
We find as the dominant pairings two degenerate nematic $p$-like solutions of the $E'$ $z$-triplet type (the singlet admixture is small), in line with the results obtained by \citeauthor{royUnconventionalPairingIsing2024}~\cite{royUnconventionalPairingIsing2024} for their model interaction. 
These solutions are followed by the SOC-introduced combinations of the $A'_1$ $s$-like singlet and $f$-like $z$-triplet that were previously discussed as candidates for the superconducting phase of monolayer NbSe$_2$~\cite{shafferCrystallineNodalTopological2020,wanObservationSuperconductingCollective2022,horholdTwobandsIsingSuperconductivity2023,shafferWeakcouplingTheoryPair2023,royUnconventionalPairingIsing2024}.
The critical temperatures of these solutions depend on the associated eigenvalues $\lambda$ of the pairing matrix and the cutoff in energy.
\begin{figure*}[th]
    \includegraphics[width=\textwidth]{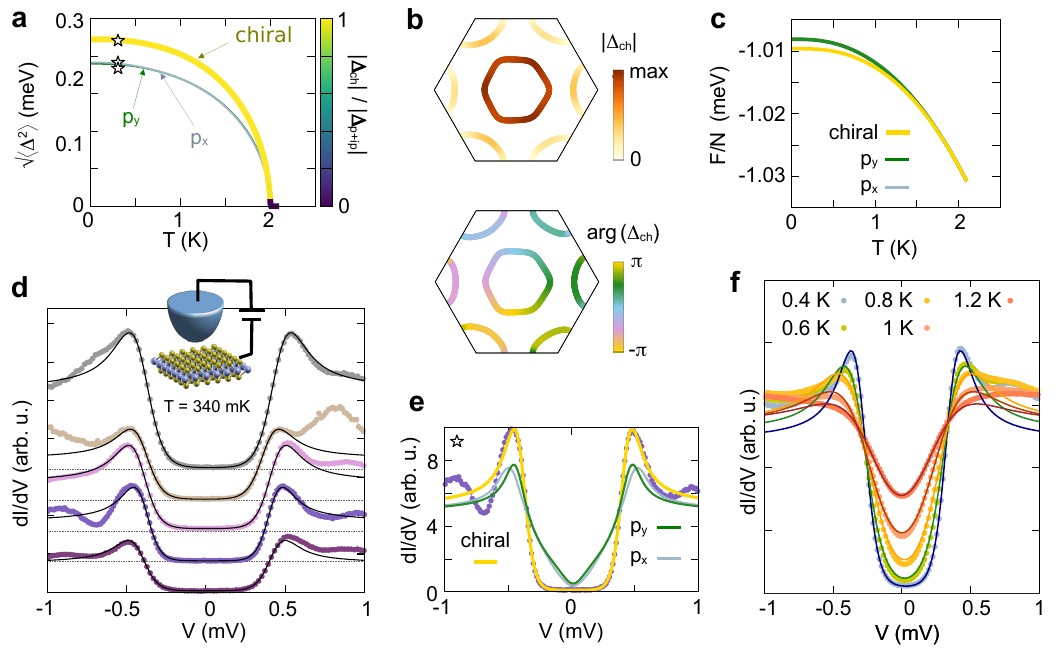}
    \caption{
        {\bf{Chiral ground state and comparison with spectroscopic measurements.}}
        $\bf{a}$ Square root of the average square amplitude  of the first three gaps solving the self-consistent gap equation as a function of temperature. The thick chiral curve is the solution with the lowest free energy; its color encodes the fraction of the chiral ($p_x+ip_y$) contribution in the leading solution which is one all the way until $T_{\T{c}}$.
        {\bf b} At $T=340$~mK the gap equation predicts a fully gapped chiral phase of $p+ip$ type.
        The phase of the order parameter winds clockwise around each contiguous Fermi surface. 
        $\bf{c}$ Free energy per electronic state  for the first three gap solutions, calculated in the range of energies $\pm2$~meV. The absence of nodes makes the chiral one  the preferred solution at low $T$. 
        $\bf{d}$ Differential conductance spectra (curves offset for clarity) recorded by STM at 340 mK, at different positions along a  NbSe$_2$ monolayer grown on a bilayer graphene/SiC(0001) substrate. A clear hard gap is observed, which can be fitted using the chiral gap in {$\bf{a}$}.    
        $\bf{e}$ One of the experimental $\T{d}I/\T{d}V$ curves in $\bf{d}$, fitted with  the gaps $\Delta_{\bk\sigma}$ obtained from the nematic and chiral solutions of the self-consistent gap equation (see the marker in $\bf{a}$). A very good agreement is found for the chiral solution on the $\Gamma$ Fermi surface; the nematic ones display a clear $V$-shape gap that departs from the experimental data.
        $\bf{f}$ Experimental traces for different temperatures and corresponding theoretical fits with the leading $\Delta_{\text{ch}}$ in $\bf{b},\bf{c}$. 
    } 
    \label{fig:chiral gap} 
\end{figure*}

Note that there is a range of eigenvalues $\lambda$ reported in the literature for corresponding solutions of model Hamiltonians~\cite{shafferCrystallineNodalTopological2020,wanObservationSuperconductingCollective2022,shafferWeakcouplingTheoryPair2023,horholdTwobandsIsingSuperconductivity2023,royUnconventionalPairingIsing2024}.
This is in line with these works using the interaction strength (and the cutoff $\Lambda$) as a fit parameter to match the critical temperature observed in experiments, e.g.  
$T_{\T{c}}\approx 2\,\T{K}$ in epitaxially grown samples~\cite{wanObservationSuperconductingCollective2022}, and in mechanically exfoliated ones~\cite{caoQualityHeterostructuresTwoDimensional2015}.
Solving the linearized gap equation for a cutoff $\Lambda=100\,\T{meV}$, we find that the pairing strength required to yield $T_{\T{c}}=2\,$K is $\lambda\approx-0.154$. 
The difference between our result and this value is well within the uncertainty of our calculation due to the differences in the reported band structures for monolayer NbSe$_2$~\cite{kimQuasiparticleEnergyBands2017,heMagneticFieldDriven2018,royUnconventionalPairingIsing2024}, and to the charge transfer from the substrate which was not included in our calculation (see also the discussion in the Methods and \cref{sm-sec: dependence on chemical potential} in the Supplementary Material).
In order to account for these uncertainties, we keep $\Lambda=100$~meV and choose to rescale instead the resulting interaction $\hat{V}\rightarrow\gamma\hat{V}$, with $\gamma \approx2.4$ to fix $T_{\T{c}}=2\,$K.
Furthermore, accounting for an additional phonon-mediated interaction~\cite{bardeenTheorySuperconductivity1957,dasElectronphononCouplingSpin2023}, e.g., by considering a momentum-independent interaction $|V_{\T{ph}}| \lesssim V^{\T{RPA}}(\bq_\Gamma)$ 
would be sufficient to explain this discrepancy, even with unscaled $\hat{V}$.
The two leading solutions of the pristine system are degenerate nematic gaps, which will remain   quasi-degenerate for weak enough breaking of the symmetry of the pristine system, e.g. by a substrate. 
This allows for the emergence of a chiral gap at temperatures below $T_{\T{c}}$~\cite{kallinChiralSuperconductors2016}. 
%
%
 %
To which extent a nodal nematic solution or the gapped chiral phase is the preferred ground state, cannot be determined at the level of the linearized gap equation used so far.
Instead, we solve the self-consistency equation, \cref{gap equation trs}, as a function of temperature.
Retaining the full momentum dependence in the gap equation renders the problem highly multi-dimensional.
However, the relevant physics is captured by expanding both the interaction and the order parameter in basis functions of the appropriate irreducible representations of the crystal symmetry group~\cite{shafferCrystallineNodalTopological2020,shafferWeakcouplingTheoryPair2023,hanisDistinguishingNodalNonunitary2024}, and solving a set of coupled equations for the expansion coefficients as a function of the temperature (see Methods).
We tracked the evolution of the coefficients down to zero temperature (see \cref{sm-subsec: expansion coefficients} of the Supplementary Material).
The square root of the average square gaps (see Methods for the definition)  
is shown in ~\cref{fig:chiral gap}a.
The magnitude and phase of this chiral gap at $T=340\,\T{mK}$ are displayed in Fig.~\ref{fig:chiral gap}b.
The free energy shown in \cref{fig:chiral gap}c for the different candidate phases confirms that the chiral gap is physically stable within the considered temperature range, cf. also the Methods.
Hence, similarly to what was possibly observed in 4Hb-TaS$_2$~\cite{silberTwocomponentNematicSuperconductivity2024}, we predict a chiral ground state also for monolayer NbSe$_2$. 
\\
A natural question is thus if this finding is compatible with available spectroscopic data.
To this end, we turn to our own low-temperature scanning tunneling spectroscopy (STS) measurements carried out on monolayer NbSe$_2$ on a bilayer graphene/SiC(0001) substrate, see~\cite{wanObservationSuperconductingCollective2022} for details of the setup. 
The measured observable is the tunneling differential conductance ($\T{d}I/\T{d}V$), which is proportional to the spectral function of the sample (see Methods and \cref{Kubo conductance formula}).
The experimental $\T{d}I/\T{d}V$ spectra recorded at different locations at $340\,\T{mK}$ are shown in Fig.~\ref{fig:chiral gap}d and fitted to the tunneling density of states (DOS) obtained from the chiral gap.
As shown in Fig. \ref{fig:bands}, in NbSe$_2$ the quasiparticle states near the $K$ and $K'$ valleys arise from the planar $d_{2,\pm 2}$ orbitals, while the ones at the $\Gamma$ valley have $d_{2,0}$ character and extend farther beyond the NbSe$_{2}$ plane.
This fact, combined with the sensitivity of STM to states with small in-plane momentum~\cite{tersoffTheoryApplicationScanning1983,ugedaCharacterizationCollectiveGround2016}, suggests a larger coupling to the states on the $\Gamma$ Fermi surface compared to those near the $K$ and $K'$ valleys.
The theoretical differential conductance shown in Fig.~\ref{fig:chiral gap}d is calculated using the coupling to the $\Gamma$ and $K$ Fermi pockets as free parameters (see Methods).

In Fig.~\ref{fig:chiral gap}e we show the fit to a representative experimental $\T{d}I/\T{d}V$ spectrum recorded at $340\,\T{mK}$, using both the chiral and the nematic solutions on the $\Gamma$ Fermi pocket. 
The presence of nodal lines in the nematic phases is reflected in a $V$-like shape of the gap, which was not observed in the experiment.
Such a shape was instead recently reported for spectroscopy experiments on monolayer 1H-TaS$_2$~\cite{vanoEvidenceNodalSuperconductivity2023}. 

Finally, the temperature evolution of the differential conductance at one fixed tip location is shown in  Fig.~\ref{fig:chiral gap}f.
The agreement with the theoretical curves which use the temperature dependence of the gap as displayed in Fig.~\ref{fig:chiral gap}a is very good.  

While one can confidently state that the STS data are consistent with the theory presented in this work, and in particular with a chiral $p$-like gap at the lowest temperatures, it is clear that they \textit{alone} are not conclusive evidence for this pairing symmetry.
In fact,  a conventional $s$-like BCS gap would yield a reasonable agreement with the data as well~\cite{wanObservationSuperconductingCollective2022}.
Additional insight is provided by nematic features observed in resistance experiments~\cite{hamillTwofoldSymmetricSuperconductivity2021,choNodalNematicSuperconducting2022}, with critical temperature dependent on the direction of the in-plane magnetic fields, in agreement with our prediction of a nematic  instability near $T_{\T{c}}$.
If one of the in-plane symmetries of $D_{\rm 3h}$ were broken, either by the application of strain or of an in-plane magnetic field, one of the nematic phases would be favored over the other, and in the  temperature range between their respective $T_{\T{c}}$ one would see the single surviving nematic phase. 
These observations suggest that the chiral phase should be thee most probable gapped phase observed in the STS experiment. 
To gather further evidence, we propose testing other signatures of a  chiral phase, which can be revealed in quasiparticle interference patterns~\cite{nagHighlyAnisotropicSuperconducting2024} or in the enhancement of the zero bias conductance near the edges of superconducting domains~\cite{mingEvidenceChiralSuperconductivity2023}.

\bibliography{references}
\clearpage

\section*{Methods}

\subsection*{Low energy model}
We consider the low-energy Hamiltonian for the metallic valence band of monolayer NbSe$_2$
\begin{align}
    \label{singleband interacting Hamiltonian}
    \hat{H}_{\T{VB}}=\sum_{\bk,\sigma} \xi_{\bk,\sigma}\opc_{\bk,\sigma}^\dagger \opc_{\bk,\sigma}+\hat{V}\,,
\end{align}
where the first term describes independent Bloch electrons in the valence band with energies $\xi_{\bk,\sigma}$, and 
$\opc_{\bk,\sigma}^{(\dagger)}$ 
are the creation and annihilation operators of Bloch electrons with crystal momentum $\bk$.
The second term is the screened Coulomb interaction which, in the position representation $(1\equiv\bm{r},\sigma)$, is defined by a Dyson equation for its matrix elements ${V(1,2)=\bra{(1,2)}\hat{V}\ket{(1,2)}}$ as
\begin{align}
    V(1,2)= V_0(1,2) +\int d(3,4) V_0(1,3)P(3,4)V(4,2)\,,
    \label{Dyson equation wrpa}
\end{align}
with $V_0(1,2)$ the matrix elements of the unscreened interaction and $P$ the polarizability of the interacting system. 
We consider a trigonal 2D lattice with periodic boundary conditions composed of $N$ prismatic unit cells of area $\Omega$ and height $L$ with sample volume $\cal{V}=N\Omega L$. 
The derivation of \cref{formal solution dyson equation in 2d} from \cref{Dyson equation wrpa} can be found in the Supplementary Material.

\subsection*{Screened interaction in RPA}
We treat the screening at the RPA level, where the full polarizability in \cref{Dyson equation wrpa} is replaced with the noninteracting polarizability tensor given by \cref{susceptibility} where
\begin{align}
    &\chi_\sigma(\bm{p}+\bm{q},\bm{p};\omega)=\frac{1}{N\Omega}\frac{f(\xi_{\bm{p}+\bq,\sigma})-f(\xi_{\bm{p},\sigma})}{\omega+\xi_{\bm{p}+\bq,\sigma}-\xi_{\bm{p},\sigma}+i\eta}\,,
    \label{susceptibility sigma}
\end{align}
$f$ is the Fermi function evaluated from the chemical potential $\mu$ and the overlaps $\Fm$ are defined in \cref{overlap factors}.
For momenta $\bk/\bk'$ outside of the first Brillouin zone, \cref{overlap factors} is modified by decomposing $\bk=\Pm(\bk)+\Qm(\bk)$ (see \cref{sm-fig: momentum decomposition} in the Supplementary Material), where $\Pm$ projects onto the reciprocal lattice, $\Qm$ projects onto the associated crystal momentum and using $\bm{u}_{\bk,\sigma}(\bm{r})=e^{i\Pm(\bk)\cdot{r}}\bm{u}_{\Qm(\bk),\sigma}(\bm{r})$.
We express the interaction $\hat{V}$ entering~\cref{singleband interacting Hamiltonian} in the Bloch basis. 
It holds for the statically screened Coulomb interaction between Bloch states (see the Supplementary Material for the full derivation),   
\begin{align}
    \label{matrix element bloch states expanded}
    &\bra{\Qm(\bk+\bq),\sigma;\Qm(\bk'-\bq),\sigma'}\hat{V}\ket{\bk,\sigma;\bk',\sigma'}=\\
    &\frac{1}{N\Omega}\sum_{\bG,\bG'}V_{\bm{G},\bm{G}'}^{\T{2D,RPA}}(\bq;0^+)\Fm^{\sigma}_{\bk+\bq,\bk}(-\bm{G})\Fm^{\sigma'}_{\bk'-\bq,\bk'}(\bm{G}')\,,\nonumber
\end{align}
where $\bk,\bk'$ and $\bq$ are in-plane momenta restricted to the first Brillouin zone.
Noting that the dominant contribution to the matrix element is due to the first few nearest neighbor sites (see \cref{sm-subsec: umklapp} of the Supplementary Material for a discussion) we introduce an effective interaction potential
\begin{align}
    \label{matrix element simplified}
    &\bra{\Qm(\bk+\bq),\sigma;\Qm(\bk'-\bq),\sigma'}\hat{V}\ket{\bk,\sigma;\bk',\sigma'}=\\
    &\approx\frac{1}{N\Omega}\left(\sum_{\bG,\bG'}V_{\bm{G},\bm{G}'}^{\T{2D,RPA}}(\bq)\right)\Fm^{\sigma}_{\bk+\bq,\bk}(\bm{0}) \Fm^{\sigma'}_{\bk'-\bq,\bk'}(\bm{0}')\nl
    &=:\frac{1}{N\Omega}V^{\T{RPA}}(\bq)\Fm^{\sigma}_{\bk+\bq,\bk}(\bm{0}) \Fm^{\sigma'}_{\bk'-\bq,\bk'}(\bm{0}')\,,\nonumber
    \end{align}
where we restrict $\bm{G},\bm{G}'$ to the range where $\Fm(\bG)\approx\Fm(\bm{0})$ holds, i.e., up to third nearest neighboring Brillouin zones.
This effective interaction potential enables a discussion of the screened interaction on the lattice in terms of a simple $\bq$-dependent quantity, but is {\em not} used for the calculation of the pairing, where we keep the dependence of  the Bloch factors $\Fm$ on $\bG$.
\subsection*{Linearized gap equation}
To access the possible superconducting instabilities described by \cref{gap equation trs}, we solve a linearized form of the gap equation.
First, we project the gap equations on the Fermi surfaces by introducing a local density of states $\rho_{\bk,\sigma}$ along the Fermi surfaces.
We restrict the allowed momenta $\bk'$ in \cref{gap equation trs} to a range around the Fermi surfaces given by a cutoff $\pm\Lambda$ in energy.
Next, we convert the sum over the transverse component $\bk'-\bk_{\T{F}}$ into an integral in energy over this range $[-\Lambda,\Lambda]$.
Assuming that the gap and the interaction vary slowly in this thin shell around each Fermi surface, we can approximate $\Delta_{\bk',\sigma}\approx\Delta_{\bk_{\T{F}},\sigma}$ and $V_{\bk,\bk',\sigma}\approx V_{\bk,\bk_{\T{F}},\sigma}$.
The resulting gap equation with $\bk,\bk'$ on the Fermi surfaces takes the form
\begin{align}
    \label{gap equation projected on FS}
    \Delta_{\bk,\sigma}&=-\sum_{\bk'\in\T{FS}} V_{\bk,\bk',\sigma}\int_{-\Lambda}^{\Lambda}d\xi\rho_{\bk',\sigma}(\xi)\Pi(E_{\bk'\sigma})\Delta_{\bk',\sigma}\,,
\end{align}
with the pairing function $\Pi(E)=\tanh(\beta E/2)/2E$. 
Close to $T_{\T{c}}$ we have $\Pi(E)\approx\Pi(\xi)$ and \cref{gap equation projected on FS} becomes a set of linear equations. 
There, the problem of finding the shape of the order parameter along the Fermi surfaces is equivalent to determining the kernel of a matrix $\mathbb{M}_\sigma$, whose indices are the momenta $\bk_{\T{F}}$ on the Fermi surface, as
\begin{align}
    \label{gap equation linearized}
    \mathbb{M}_{\sigma}\Vec{\Delta}_{\sigma}=0\,,
 \end{align}   
    where, $\forall \bk,\bk'\in\T{FS}$,  
    \begin{align}
    (\mathbb{M}_\sigma)_{\bk,\bk'}&=\delta_{\bk,\bk'}+V_{\bk,\bk',\sigma}\rho_{\bk',\sigma}\alpha^0(T,\Lambda)\,,\\
    \label{pairing function linearized}
    \alpha^0(T,\Lambda)&=\int_{0}^{\Lambda}d\xi \frac{\tanh(\beta\xi/2)}{\xi}\,.
\end{align}
Here, we assumed a flat density of states $\rho_{\bk,\up}(\xi)\approx\rho_{\bk,\up}$ within the energy range $\pm \Lambda$ around the chemical potential.
We rewrite $\mathbb{M}_\sigma$ by factoring out the energy integral $\alpha^0$ 
 in the second summand as
\begin{align}
    \label{definition U matrix}
    \mathbb{M}_\sigma=\mathbbm{1} +\alpha^0(T,\Lambda) \mathbb{U}_{\sigma}\,,
\end{align}
where $\mathbb{U}_\sigma$ is the pairing matrix containing the interaction and local density of states along the Fermi surfaces.
The possible superconducting pairings solving \cref{gap equation linearized} are the eigenvectors of $\mathbb{U}_{\up}$ with the corresponding eigenvalues $\lambda_i$ fulfilling 
\begin{align}
   \lambda_i=-1/\alpha^0(T_{\T{c},i},\Lambda) \,,
\end{align}
and $T_{\T{c},i}$ the critical temperature of the respective instability.
The cutoff $\Lambda$ sets an energy scale against which to measure the gaps and $T_{\T{c}}$, similar to the Debye frequency in the usual BCS-theory~\cite{bardeenTheorySuperconductivity1957}.
In general, not all eigenvectors of $\mathbb{U}_\sigma$ found in this way correspond to physically stable situations~\cite{hutchinsonMixedTemperaturedependentOrder2021}.
Similarly, the requirement of vanishing $\Delta_{\bk,\sigma}$ close to $T_{\T{c},i}$ is only fulfilled if no other superconducting phase with higher $T_{\T{c}}$ is present.
As such, we expect only the solution with the highest $T_{\T{c}}$ to be realized.
Due to the monotonic nature of $\alpha^0(T,\Lambda)$ as a function of both $T$ and $\Lambda$, the solution with the highest critical temperature always corresponds to the lowest negative eigenvalue $\lambda$ of the pairing matrix.
As $\Lambda$ is a model parameter, we classify and discuss the strength of the different instabilities by their eigenvalues $\lambda_i$.

\subsection*{Expansion of the temperature-dependent gap equation in basis functions}
To investigate how the gaps evolve at temperatures below $T_{\T{c}}$, we solve \cref{gap equation projected on FS} without linearizing.
Since we want to use the approximations $\Delta_{\bk',\sigma}\approx\Delta_{\bk_{\T{F}},\sigma}$ and $V_{\bk,\bk',\sigma}\approx V_{\bk,\bk_{\T{F}},\sigma}$, we fix the cutoff to the  value of $\Lambda=0.1\,$eV which, together with the rescaling of the screened interaction to $\hat{V}_{\gamma}:=\gamma\hat{V}$ as discussed in the main text, yields $T_{\T{c}}=2$~K.
This enables us to fit to the experimental $T_{\T{c}}$ while retaining the shape of the interaction as found from our calculation of the screened Coulomb interaction.
Since the macroscopic polarizability $P_{\bf{0,0}}$ calculated from DFT (cf.~\cref{sm-subsec: dft} of the Supplementary Material) is even lower at intermediate momentum transfers than our results from a tight-binding calculation, it is plausible that the screened interaction in absence of the substrate is even stronger than our $\hat{V}$.
This is to be expected, since the inclusion of the Se $p$-orbitals which contribute most noticeably to the valence bands near the M points, would further diminish the overlaps between states on the Fermi surface, resulting in lower screening.

We expand the gap for one spin species in terms of a restricted set of basis functions, corresponding to the irreducible representations of the symmetry group $D_{\T{3h}}$, as 
\begin{align}
    \label{expansion gap}
    \forall_{\bk\in\pi}:\quad \Delta_{\bk,\uparrow}=\sum_{\mu}f^\pi_\mu(\bk)\Delta^\pi_\mu\,,
\end{align}
where $\mu$ labels basis functions and $\pi\in\{\Gamma,K\}$  the Fermi surfaces~\cite{roldanInteractionsSuperconductivityHeavily2013,shafferCrystallineNodalTopological2020}.
We orthonormalize our basis functions with respect to the number $N_{\pi}$ of momenta on each Fermi surface $\pi$ to satisfy
\begin{align}
    \label{orthogonaity basis functions}
    \langle f^\pi_\mu,f^\pi_{\mu'}\rangle_\pi:=\frac{1}{N_{\pi}}\sum_{\bk\in\pi}f^{\pi}_{\mu'}(\bk)f^{\pi}_{\mu}(\bk)=\delta_{\mu,\mu'}\,.
\end{align}
Using this orthogonality, we reduce the dimensionality of the self-consistency problem drastically by projecting the gap equation into a form that couples only the expansion coefficients $\Delta^{\pi}_{\mu}$~\cite{shafferCrystallineNodalTopological2020}.
It reads
\begin{align}
    \label{projected out gap equation}
    \Delta^\pi_\mu=-\frac{1}{N_{\pi}}\sum_{\pi',\mu'}\sum_{\substack{\bk\in\pi\\ \bk'\in\pi'}}f_\mu^\pi(\bk)V_{\bk,\bk',\uparrow}\;\rho_{\bk',\uparrow}\;\alpha^{\pi'}_{\bk'}\;f^{\pi'}_{\mu'}(\bk')\Delta^{\pi'}_{\mu'}\,, 
\end{align}
where the pairing strength $\alpha^\pi_{\bk}$ is given by
\begin{align}
    \label{pairing strength}
    \alpha^{\pi}_{\bk}=\int_{0}^\Lambda d\xi\,\frac{\tanh(\beta E(\xi,\Delta^{\pi}_{\bk}) /2)}{E(\xi,\Delta^{\pi}_{\bk})}\,.
\end{align}
At $T_{\T{c}}$, the pairing strength $\alpha^\pi_{\bk}$ becomes independent of $\bk$ and $\pi$.
In this case, the orthogonality \cref{orthogonaity basis functions} yields a linearized gap equation for the expansion coefficients 
\begin{align}
    \label{gap equation basis functions linearized}
    \Delta^\pi_\mu=-\alpha^0 \sum_{\pi',\mu'} U^{\pi,\pi'}_{\mu,\mu'} \Delta^{\pi'}_{\mu}\,,
\end{align}
where the pairing matrix $\mathbb{U}$ in this representation is given by
\begin{align}
    U^{\pi,\pi'}_{\mu,\mu'}&=\frac{1}{N_{\pi}}\sum_{\bk\in\pi,\bk'\in\pi'}f^{\pi}_{\mu}(\bk)V_{\bk,\bk',\uparrow}\;\rho_{\bk',\uparrow}f^{\pi'}_{\mu'}(\bk')\,.
\end{align}
To account for the mixing between different symmetries due to the $\bk$ dependence of $\alpha^\pi_{\bk}$ below $T_{\T{c}}$, we insert a unity in \cref{projected out gap equation}
to arrive at 
\begin{align}
    \label{gap equation basis functions}
    \Delta^\pi_\mu=-\sum_{\pi',\mu',\mu''} U^{\pi,\pi'}_{\mu,\mu'}\Upsilon^{\pi'}_{\mu',\mu''} \Delta^{\pi'}_{\mu''}\,,
\end{align}
where the mixing of different basis functions is described by
\begin{align}
    \label{definition upsilon}
    \Upsilon^{\pi}_{\mu,\mu'}&=\frac{1}{N_{\pi}}\sum_{\bk\in\pi} f^{\pi}_{\mu}(\bk)\alpha^\pi_{\bk}f^{\pi}_{\mu'}(\bk)\,.
\end{align}
For an incomplete set of basis functions this insertion of unity becomes approximate, but in our case still suffices to capture the weak mixing between basis functions induced by $\alpha^\pi_{\bk}$.
As such, we employ \cref{gap equation basis functions} to numerically determine the evolution of the solutions obtained from the linearized equations below their $T_{\T{c},i}$.
The basis functions we used and the temperature dependence curves for the respective expansion coefficients are shown in \cref{sm-subsec: expansion coefficients} of the Supplementary Material.

We define a pure chiral solution as
\begin{equation}
  \Delta_\text{p+ip} = \frac{||\Delta_{\T{py}}||}{||\Delta_{\T{px}}||} \Delta_{\T{px}} + i\Delta_{\T{py}},  
\end{equation}
where  $||\Delta||:=\sqrt{\sum_{\pi=G,K}\langle\Delta,\Delta\rangle_\pi}$, with the scalar pro-duct defined in \cref{orthogonaity basis functions}.
The $\Delta_{\T{px}}$ component is renorm\-alized so that both nematic solutions enter with the same weight. 
The admixture of $\Delta_\text{p+ip}$ in the chiral solution $\Delta_\text{ch}$ in \cref{fig:chiral gap}c is defined as
\begin{equation}
 \frac{\Delta_\text{p+ip}}{\Delta_{\text{ch}}}  := \frac{ |\sum_{\pi=\Gamma,K}\langle \Delta_\text{p+ip}, \Delta_{\mathrm{ch}}\rangle_\pi|}{||\Delta_\mathrm{ch}||\, ||\Delta_{\mathrm{p+ip}}||}\,.
\end{equation}
The square root of mean squared $\Delta$ shown in \cref{fig:chiral gap}a for each gap symmetry is $\sqrt{\langle \Delta^2\rangle} \equiv ||\Delta||$.
\subsection{Free energy for self-consistent solutions}
The free energy of a superconductor with gap $\Delta_{\bk\sigma}$ can be written as
\begin{align}
    \label{free energy}
    F =&  \frac{1}{2}\sum_{\bk\sigma} \left( \xi_{\bk\sigma} - E_{\bk\sigma} + \Pi(E_{\bk\sigma}) |\Delta_{\bk\sigma}|^2\right) \nonumber \\
    & - k_B T \sum_{\bk\sigma} \ln\left( 1 + e^{-E_{\bk\sigma}/k_BT}\right).
\end{align}
Note that this free energy is written assuming that the gap obeys the self-consistent equations (see \cref{sm-model for sc in nbse2} of the Supplementary Material comments on its derivation).
It cannot be used to find the physical gaps by minimization with respect to $\Delta_{\bk\sigma}$, but does yield the correct result once a solution is found by solving the self-consistent gap equations~\cite{hutchinsonMixedTemperaturedependentOrder2021}.
We therefore use it to rank the solutions by their free energy gain within the energy window of $\pm 2$~meV where the contributions of different gap symmetries are distinguishable.
\subsection*{Experimental setup and methods}
NbSe$_2$ monolayers$_2$ were grown on graphitized (bilayer graphene) on 6H-SiC(0001) by molecular beam epitaxy (MBE) at a base pressure of $\sim2\cdot10^{-10}\,\T{mbar}$ in our home-made ultrahigh-vacuum (UHV) MBE system.
SiC wafers with resistivities $\rho\sim 120\,\Omega\T{cm}$ were used~\cite{rubio-verduVisualizationMultifractalSuperconductivity2020}.
\\
Reflective high-energy electron diffraction (RHEED) was used to monitor the growth of the NbSe layer$_2$. During the growth, the bilayer graphene/SiC substrate was kept at $580^\circ\,\T{C}$.
High-purity Nb (99.99\%) and Se (99.999\%) were evaporated using an electron beam evaporator and a standard Knudsen cell, respectively.
The Nb:Se flux ratio was kept at 1:30, while evaporating Se led to a pressure of $\sim 5\cdot 10^{-9}\,\T{mbar}$ (Se atmosphere).
Samples were prepared using an evaporation time of $30\,\T{min}$ to obtain a coverage of $\sim 1$ ML.
To minimize the presence of atomic defects, the evaporation of Se was subsequently kept for an additional 5 minutes.
Atomic force microscopy (AFM) under ambient conditions was routinely used to optimize the morphology of the NbSe$_2$ layers.
The samples used for AFM characterization were not further used for the scanning tunneling microscope (STM).
Lastly, to transfer the samples from our UHV-MBE chamber to our STM, they were capped with a $\sim 10\,\T{nm}$ amorphous film of Se, which was subsequently removed by annealing at the STM under UHV conditions at 280 ° C.

Scanning tunneling microscopy/spectroscopy 
experiments were carried out in a commercial low-temperature, high magnetic field (11 T) UHV-STM from Unisoku (model USM1300) operated at $T=0.34\,\T{K}$ unless otherwise stated.
STS measurements (using Pt/Ir tips) were performed using the lock-in technique with typical ac modulations of $20-40\,\T{µV}$ at $833\,\T{Hz}$. All data were analyzed and rendered using WSxM software~\cite{horcasWSXMSoftwareScanning2007}.
To avoid tip artifacts in our STS measurements, the STM tips used for our experiments were previously calibrated on a Cu(111) surface.
A tip was considered calibrated only when tunneling spectroscopy performed on Cu(111) showed a sharp surface state onset at $-0.44\,\T{eV}$ followed by a clean and monotonic decay of the differential conductance signal (i.e. $\T{d}I/\T{d}V$).
In addition to this, we also inspected the differential conductance within $\pm 10\,\T{mV}$ to avoid strong variations around the Fermi energy.

\subsection*{Comparison to experimental tunneling spectroscopy}
For spectroscopy in the tunneling regime with a normal metal tip, the differential conductance is given by
\begin{align}
    \label{Kubo conductance formula}
    G(V)=\sum_{\bk, \sigma} C_{\bk \sigma}\int_{-\infty}^{\infty}\frac{\T{d}E}{2\pi}\;A_{\T{s}}(\bk,\sigma;E)\left(-\frac{\partial f(E+eV)}{\partial E}\right)\,,
\end{align}
where $A_{\T{s}}$ is the spectral function of the superconductor.
The coupling $C_{\bk \sigma}$ accounts for the spectral function of the tip and the tunneling overlap between the tip's evanescent states and quasiparticle states from the superconductor with momentum $\bk$ and spin $\sigma$.
We do not expect the coupling to vary much on any given Fermi surface $\pi$, and as such we approximate $C_{\bk,\sigma}\approx C_{\pi}$.
However, due to the different orbital composition of the quasiparticle states at the  $K$, $K'$ valleys and at the $\Gamma$ surface, cf. Fig.~\ref{fig:bands}b, the coupling constant will be in general different.
In particular, since the $d_{2,0}$ orbital extends farthest out of plane and the STM favors states with small in-plane momentum, we expect $C_{\Gamma}\gg C_{K}$. 

In the tunneling regime, where the experimental data was taken, the lifetime of the quasiparticles is weakly affected by the coupling to the tip and the spectral function can be approximated with the local density of states of the quasiparticles,  $A_s(\bk,\sigma,E)=2\pi D_{\bk,\sigma}(E)$.
The tunneling to each of the Fermi surfaces represents a distinct transport channel, whose strength is given by the sum over the local  density of states
\begin{align}
    \label{angleDependentDOS}
    D_{\pi,\sigma}(E)=\sum_{\bk\in\pi}D_{\bk,\sigma}(E)\,.
\end{align}
The 
latter are modeled by assuming a BCS form factor modifying the local density of states $\rho_{\bk,\sigma}$ in the normal conducting state as
\begin{align}
    \label{BCS form factor}
    D_{\bk,\sigma}(E)=\rho_{\bk,\sigma}\T{Re}\sqrt{\frac{E^2}{E^2-|\Delta_{\bk,\sigma}|^2}}\,.
\end{align}
In the absence of magnetic fields, we can assume that, due to time reversal symmetry, $D_{\pi,\sigma}=D_{\pi}/2$ and $C_{\pi,\sigma}=C_{\pi}$.
As such, \cref{Kubo conductance formula} predicts for the total differential conductance a superposition of contributions from the distinct transport channels with weights provided by the couplings $C_{\pi}$ as
\begin{align}
    \label{fit function}
    G(V)&\approx\sum_{\pi}C_{\pi}G_\pi(V)\,,\\
    G_\pi(V)&=\int_{-\infty}^{\infty}\T{d}E\;D_{\pi}(E)\left(-\frac{\partial f(E+eV)}{\partial E}\right).
    \label{differential conductance per transport channel channel}
\end{align}
For low temperature, the form of the gap on each separate Fermi surface results in characteristic signatures in the differential conductance of the respective transport channel.
We use as a fit function this superposition of the differential conductance of the individual transport channels. 
For the gaps on the Fermi surfaces $\Delta_{\bk,\sigma}$, we use the form of the gaps as found from \cref{gap equation basis functions} at the experimental temperature $T_{\T{exp}}$
and allow for a fit of the amplitude by a common rescaling of all $\Delta_{\bk,\sigma}$ by a parameter $A$ as $\Delta_{\bk,\sigma}\rightarrow A\Delta_{\bk,\sigma}$. 
The latter is introduced to take care of the uncertainty of the actual experimentally realized $T_{\T{c}}$ and the rescaling factor $\gamma$ of the interaction.
\begin{table}[ht]
    \centering
    \begin{tabular}{c|c|c|c|c|c}
        Sol.  & $C_{K},C_{\Gamma}$\,[nS\,eV\,\AA$^2$] & $A$ & $V_0$\,[mV]& $G_0$\,[nS] & $\T{T}_{\T{eff}}$/T \\
        \hline
        ch  & 3.84\,E-6 , 8.20\,E0 & 1.17\,E0 & 1.26\,E-2 & 5.89\,E-3 & 1.31\\
        \hline
        $p_x$ & 0.00\,E-9 , 7.93\,E0 &  1.08\,E0 & 1.22\,E-2 & 0.00\,E-9 & 1.01\\
        \hline
        $p_y$ & 0.00\,E-9 , 7.90\,E0 & 1.17\,E0 & 1.29\,E-2 & 0.00\,E-9 & 1.01 \\
    \end{tabular}
    \caption{
    Coefficients of best fit between the theoretical prediction of the differential conductance in the possible superconducting phases and the experiment as shown in \cref{fig:chiral gap}\,(e).
    The best fit for each solution is achieved by selective coupling to the $\Gamma$ pocket in line with the arguments presented in the main text.
    }
    \label{table fit coefficients}
\end{table}
To qualitatively account for additional sources of broadening in the experiment, we fit with an effective temperature $T_{\T{eff}}$ in the calculation of the derivative in \cref{differential conductance per transport channel channel} which ranges between 1.3 and 1.4 times the recorded base temperature of the STM in \cref{fig:chiral gap}\,(d,e) and is kept at 1.4 times the measurement temperature for \cref{fig:chiral gap}\,(f).
To account for offsets in the calibration, we further allow for both a small constant offset $G_0$ in the measured conductivity and $V_0$ in the recorded voltage.
The fit coefficients for the other traces shown in \cref{fig:chiral gap}\,(d,f) only differ by small quantitative changes from the ones reported in \cref{table fit coefficients} for \cref{fig:chiral gap}\,(e). 
For the fitting we use the trust region reflective algorithm as implemented in SciPy's "curve\_fit" routine.
We consider  a range of $\pm 0.5\,\T{mV}$ containing the main coherence peaks, but not the satellite features due to Leggett modes~\cite{wanObservationSuperconductingCollective2022}, which are not accounted for in our theory.
The resulting fit parameters are listed in \cref{table fit coefficients}.
The best fits for the solutions associated with the $E'$ irreducible representation are obtained by considering strongly selective coupling of the tip to the $\Gamma$ pockets.
\\

\noindent{\bf Data availability}

All data shown in the figures and underpinning the shown results will be made available upon reasonable request to the authors.
The code used to generate the shown data is available upon request.\\

\noindent{\bf Acknowledgments}

The work of Julian Siegl and Anton Bleibaum was funded by project B09 within CRC 1277 and RTG {2905 - 502572516} of the Deutsche-Forschungsgemeinschaft, respectively.
M.K. appreciates discussions with Timon Mo\v{s}ko  and acknowledges support from the Interdisciplinary Centre for Mathematical and Computational Modelling (ICM), University of Warsaw (UW), within grant no. G83-27 and the financial support from the National Center for Research and Development (NCBR) under the V4-Japan project BGapEng V4-JAPAN/2/46/BGapEng/2022.
M.M.U. acknowledges support by the ERC Starting grant LINKSPM (Grant \#758558) and by the grant PID2023-153277NB-I00 funded by the Spanish Ministry of Science, Innovation and Universities. 
M.M. appreciates helpful discussions with Fernando de Juan, Martin Gmitra and Cosimo Gorini, and acknowledges thankfully the help of Andrea Donarini and Marko Milivojevic in constructing the basis functions. 
J.Si. appreciates discussions with Rui Shi and acknowledges his help by pointing out a mistake in a previous version of this manuscript.\\

\noindent{\bf Author contributions}

J.Si. and M.M. performed the analytical and numerical tight-binding calculations.
M.K. calculated the DFT band structure and polarizability.
W.W. and M.M.U. performed the STM/STS experiments on the monolayer NbSe$_2$.
A.B. performed the fitting to the experimental data.
M.G. conceived the project, leading with M.M. the theoretical analysis with support from J.Si. and J.Sc.; M.M.U. was responsible for the experimental part of this work.  
The manuscript was written by M.G., J.Si. and M.M. with contributions from all other authors. 
\makeatletter\@input{supp2main.tex}\makeatother
\end{document}


\date{\today}
\title{Supplementary material for: Friedel oscillations and chiral superconductivity in monolayer NbSe\texorpdfstring{$_2$}{2}}
\author{Julian Siegl}
\email{Julian.Siegl@ur.de}
\author{Anton Bleibaum}
\affiliation{Institute for Theoretical Physics, University of Regensburg, 93 053 Regensburg, Germany}
\author{Wen Wan}
\affiliation{Donostia International Physics Center, Paseo Manuel de Lardizábal 4, 20018 San Sebastián, Spain.}
\author{Marcin Kurpas}
\affiliation{Institute of Physics, University of Silesia in Katowice, 41-500 Chorzów, Poland}
\author{\hbox{John Schliemann}}
\affiliation{Institute for Theoretical Physics, University of Regensburg, 93 053 Regensburg, Germany}
\author{Miguel M. Ugeda}
\affiliation{Ikerbasque, Basque Foundation for Science, Bilbao 48013, Spain}
\affiliation{(CSIC-UPV-EHU), Paseo Manuel de Lardizábal 5, San Sebastián 20018, Spain}
\author{Magdalena Marganska}
\affiliation{Institute for Theoretical Physics, University of Regensburg, 93 053 Regensburg, Germany}
\affiliation{Department of Theoretical of Physics, Wrocław University of Science and Technology, Wybrzeże Wyspiańskiego 27, 50-370 Wrocław, Poland}
\author{Milena Grifoni}
\email{Milena.Grifoni@ur.de}
\affiliation{Institute for Theoretical Physics, University of Regensburg, 93 053 Regensburg, Germany}

\maketitle

\tableofcontents

\section{Band structure\label{sec: bandstructure}}
The starting point for the calculation of both the screened interaction and the superconductivity in the main text is a model for the underlying electronic band structure $\xi_{\bk,\sigma}$ (cf. \cref{main-singleband interacting Hamiltonian}) of free-standing monolayer NbSe$_2$.

\subsection{Tight-binding model for the band structure of \texorpdfstring{NbSe$_2$}{NbSe2}\label{tight binding model}}
The screened potential was calculated employing Bloch states obtained using the tight-binding model for monolayer TMDs presented by \citeauthor{liuThreebandTightbindingModel2013}~\cite{liuThreebandTightbindingModel2013}.
Therein, the Wannier states are approximated through linear combinations of the Nb $d$-shell atomic orbitals $d_{x^2-y^2},d_{xy}$ and $d_{z^2}$.
This approach yields a realistic band structure for the valence band dominating the low-energy physics of monolayer (ML) NbSe$_2$ as well as a good approximation to the first two Nb-dominated conduction bands.
The role of the Se atoms in this model is relegated to a \hbox{modification} of the hopping within the niobium sublattice.

Multiple parametrizations using this model for monolayer NbSe$_2$ are available in the literature~\cite{kimQuasiparticleEnergyBands2017,heMagneticFieldDriven2018}.
For the data underpinning the figures in the main text, we employed the parameter set by \citeauthor{heMagneticFieldDriven2018}~\cite{heMagneticFieldDriven2018}. Since the lattice constant was not specified in that work, we used $a=3.445$~\AA ~from \cite{kimQuasiparticleEnergyBands2017}.

\subsection{First principles calculations}\label{subsec: dft}
\begin{figure}[ht]
\includegraphics[width=\columnwidth]{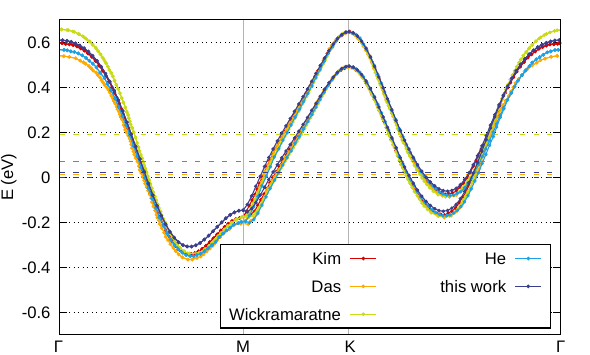}
\caption{\label{fig: DFT comparison}
Spin-split valence band of NbSe$_2$ from the DFT results in a set of selected publications.
The plot contains digitized data from Kim et al.~\cite{kimQuasiparticleEnergyBands2017}, Das et al.~\cite{dasElectronphononCouplingSpin2023}, Wickramaratne et al.~\cite{wickramaratneIsingSuperconductivityMagnetism2020}, He et al.~\cite{heMagneticFieldDriven2018} and the results of our own calculations.
Where necessary, the ${\Gamma}M$, ${\Gamma}K$ and $MK$ distances have been stretched and/or inverted to allow us to compare all data.
Each set of bands has been shifted in energy so that the top of the $K$ valleys aligns to that in [\onlinecite{kimQuasiparticleEnergyBands2017}]; the Fermi level for the shifted sets is indicated by a dashed line of the same color.
}
\end{figure}

In order to estimate how much our results for the screened potential are affected by the contributions from the Se orbitals, we calculated the screened potential also with  density functional theory (DFT). In addition, we evaluated the electronic band structure and compared it  with already existing ones.   
The different ab-initio-based band structures available in literature~\cite{kimQuasiparticleEnergyBands2017,heMagneticFieldDriven2018,wickramaratneIsingSuperconductivityMagnetism2020,dasElectronphononCouplingSpin2023} for the valence band of freestanding monolayer NbSe$_2$ are shown in Fig.~\ref{fig: DFT comparison}. 
The main differences are the height of the $\Gamma$ pocket and the position of the Fermi level within the valence band.  
For our study of superconducting pairing the latter is more important, since the electronic properties depend strongly on the shape and position of the Fermi surface.
Our first-principles calculations indicate that these differences are mainly related to the details of the crystal structure.
For smaller lattice constant the maximum of the $\Gamma$ pocket and the band edge along the $\Gamma$M path are lower in energy, while the pocket at the $K$ point remains almost unchanged, for small (less than 1\%) variation of the lattice constant.
Since small differences in calculation details, such as force convergence criteria or the chosen lattice parameters, can affect the electronic properties, the discrepancies in the band structure shown in Fig.~\ref{fig: DFT comparison} can be interpreted as parameter tolerances for the tight-binding model.

The optimization of the lattice constant and atomic positions has been done using the {\sc Quantum Espresso}~\cite{giannozziQUANTUMESPRESSOmodular2009,giannozziAdvancedcapabilitiesmaterials2017} software package.
The norm-conserving Vanderbilt pseudopotentials~\cite{hamannOptimizednormconservingVanderbilt2013,hamannErratumOptimizednormconserving2017} with the Perdew-Burke-Ernzerhof (PBE)~\cite{perdewGeneralizedGradientApproximation1996,perdewGeneralizedGradientApproximation1997} implementation of the generalized gradient approximation (GGA)  exchange-correlation potential were used.
The kinetic energy cutoffs of the plane wave basis sets were $56\,\T{Ry}$ for the wave function and $224\,\T{Ry}$ for the charge density.
The irreducible Brillouin zone was sampled with 37 $k$-points generated by the Monkhorst-Pack \cite{monkhorstSpecialpointsBrillouinzone1976,packSpecialpointsBrillouinzone1977} scheme from the $18\times18\times1$ k-point mesh.
Periodic copies of a 2D NbSe$_2$ film were separated by a vacuum of $20\,\T{\AA}$ in the $z$ direction to minimize possible spurious interaction effects.
Assuming the force and total energy thresholds $10^{-4}\,\T{Ry}$ and $10^{-5}\,\T{Ry}$, respectively, our relaxation calculations give the lattice constant $a=3.47\,\T{\AA}$, in agreement with values reported in the literature~\cite{zhengFirstprinciplesStudyCharge2018,kimQuasiparticleEnergyBands2017}. 
The optimized structure was transferred to the full-potential (linearized) augmented plane-wave ((L)APW) code Wien2K \cite{blahaWIEN2kAPW+loprogram2020}, which was used to calculate the electronic properties.
Residual internal forces were relaxed below $0.5\,\T{mRy/bohr}$.
Self-consistent calculations were done with the $30\times30\times1$ Monkhorst-Pack k-point mesh, taking the parameter $R_{MT}\cdot K_{max}$=8, where $R_{MT}$ is the muffin-tin radius and $K_{max}$ is the plane-wave cutoff.
The macroscopic electronic polarizability $P(\bq)$ was calculated using the formula
\begin{align}\label{eq:susc_dft}
    P_{\bf{0},\bf{0}}(\bq)& = \\
    \sum_{n,n'}&\sum_{\bk} \frac{f(\epsilon_{n,\bk})-f(\epsilon_{n', \bk+\bq})}
    {\epsilon_{n, \bk}-\epsilon_{n', \bk+\bq}+ i0^+}|
    \langle n \bk | e^{i \bq \cdot \br}|n'\bk+\bq \rangle|^2\,,\nonumber
\end{align}
where $n$, $n'$ are band indices and summing over 3600 k-points uniformly distributed in the entire first Brillouin zone.
For the valence band, $n=\sigma$ and \cref{eq:susc_dft} reduces to \cref{main-susceptibility}.
As the other bands are far from the Fermi energy, their contribution is a small quantitative correction to the susceptibility obtained by considering only the valence band~\cite{johannesFermisurfaceNestingOrigin2006}.

\begin{figure}[ht]
\includegraphics[width=\columnwidth]{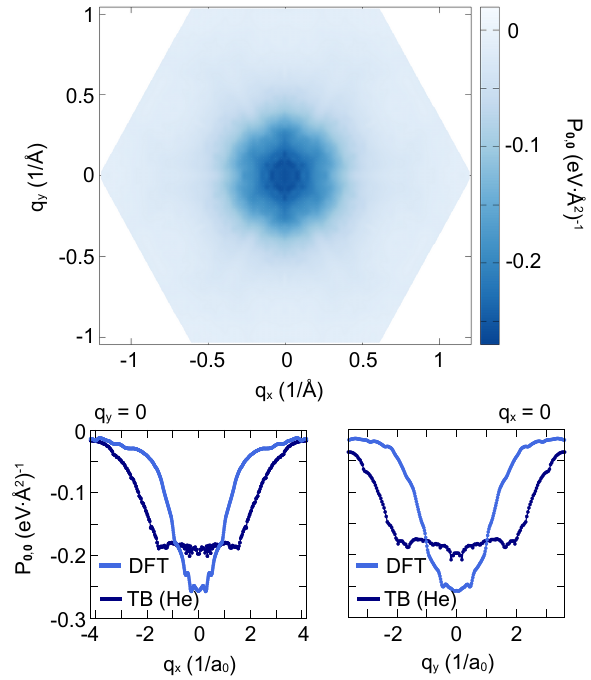}
\caption{\label{fig: susc DFT}
Macroscopic polarizability calculated from DFT, interpolated from a 60x60 to a 808x700 grid and symmetrized, and its cuts along $q_y=0$ and $q_x=0$, compared to those obtained in the TB model in the main text. The screening of low $\bq$ processes is weaker in the TB model than in DFT, due to a lower density of states, but that of $\bq\sim\bq_K$ is reduced even more  in the DFT than in the TB calculation, favoring superconducting pairing at higher $T_\T{c}$ than in the main text.
}
\end{figure}

The polarizability calculated directly from the DFT wave functions is qualitatively similar to that obtained from a tight-binding calculation (TB), see \cref{fig: susc DFT}, in that the screening is strongest for small momentum transfers. 
The screening at momenta between $\bq_\Gamma$ and $\bq_K$ is weaker in the DFT than in the TB calculation, which could be caused by the lack of the Se orbitals in the TB Hamiltonian.
The presence of these additional degrees of freedom in the Bloch spinors $ \boldsymbol{u}_{\bk\sigma}$ calculated in DFT decreases the mean overlap with $\boldsymbol{u}_{\bk+\bq,\sigma}$ for sufficiently large $\bq$.
This decreased overlap in turn suppresses $|P_{\bf{0},\bf{0}}(\bq)|$ calculated in DFT for intermediate $\bq$, resulting in weaker screening.
Due to the numerical constraints on using DFT throughout the whole calculation, we work with the band structure obtained from the tight-binding model for the calculation of $V^{\T{RPA}}$.
To account for the resulting overestimation in screening as compared to the DFT, as well as the uncertainty in the band structure, we introduce in the Methods a parameter $\gamma$ rescaling the interaction in order to fit to the experimental $T_{\T{c}}$. 

\section{Screened Coulomb interaction\label{sec: screened couloumb}}
\begin{figure}[ht]
\begin{center}
    \includegraphics[width=\columnwidth]{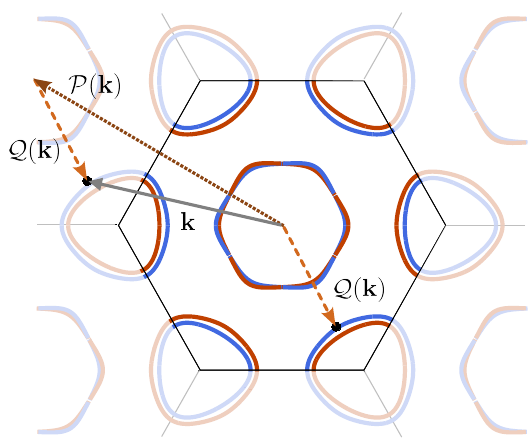}
    \caption{\label{fig: momentum decomposition}
    Decomposition of the momentum $\bm{k}$ into a reciprocal lattice momentum $\Pm(\bm{k})$ and crystal momentum $\Qm(\bm{k})$.
    The Fermi surfaces are drawn for orientation within the Brillouin zone and to highlight the periodicity of the reciprocal lattice.
    }
\end{center}
\end{figure}

The interaction in the low-energy Hamiltonian used in the main text is the screened Coulomb interaction 
\begin{align}
    \label{screened coulomb interaction hamiltonian}
    \hat{V}=\frac{1}{2}\sum_{\bk,\bk',\bm{q},\sigma,\sigma'} &V(\bk,\bk',\bm{q},\sigma,\sigma')\nl
    &\opc_{\Qm(\bk+\bm{q}),\sigma}^\dagger \opc_{\Qm(\bk'-\bm{q}),\sigma'}^\dagger \opc_{\bk',\sigma'} \opc_{\bk,\sigma}\,.
\end{align}
For convenience in treating Umklapp scattering later on, we defined the projectors $\Pm$ and $\Qm:=1-\Pm$, where $\Pm$ projects any momentum onto the closest reciprocal lattice vector.
By definition, $\Qm$ projects any momentum onto the corresponding crystal momentum inside the first BZ.
The relation between $\bm{k},\Pm(\bm{k})$ and $\Qm(\bm{k})$ is illustrated in Fig.~\ref{fig: momentum decomposition}.

Due to the metallic nature of the valence band of monolayer NbSe$_2$, this electronic interaction can be significantly altered by screening effects not just of the dielectric environment, but also the material itself.
For the case of a free-standing sample, as discussed in the main text, we expect this intrinsic screening to play a key role in enabling superconducting pairing by selectively favoring short-range (high momentum transfer) scattering processes~\cite{horholdTwobandsIsingSuperconductivity2023}.
The phenomenology of screening at long ranges in two dimensional materials can be captured by considering the charges as tightly bound to the plane of the material~\cite{cudazzoDielectricScreeningTwodimensional2011}.
However, for crystalline materials the three-dimensional nature of the orbitals involved in the scattering processes becomes important when considering the corrections due to the lattice~\cite{ceaBandStructureInsulating2020}.
In this section, starting from the three-dimensional form of the screened potential, we obtain an effective interaction on a two-dimensional lattice in the random phase approximation.
This allows us to derive an expression for the statically screened Coulomb interaction  ${\hat{V}^{\T{RPA}}= \lim_{\omega\rightarrow0^+}\hat{V}^{\T{RPA}}(\omega)}$ between Bloch states in 2D materials.

In the following $V_0(\bpos-\bpos')$ denotes the unscreened Coulomb potential and we assume periodic boundary conditions both in the plane as well as in the direction perpendicular to it.
The volume for the considered space is $\mathcal{V}=N\mathcal{V}^p$.
Here $N$ is the number of unit cells and $\mathcal{V}^p=L\Omega$ is the volume of the primitive unit cell given by the product of the in-plane area $\Omega$ and the out-of-plane periodicity $L$.
The Bloch spinors are normalized to unity over a primitive unit cell.
To recover the free-standing case, we send $L\rightarrow\infty$ in the end.

\subsection{RPA from Dyson equation in real space}

The Dyson equation \cref{main-Dyson equation wrpa} describes the effect of screening due to the noninteracting polarizability $P(3,4;\omega)$~\cite{hybertsenElectronCorrelationSemiconductors1986,hybertsenInitioStaticDielectric1987}.
To evaluate this Dyson equation in the Bloch eigenbasis of the noninteracting valence band, we start with a matrix element in this basis of a generic, in general not translationally invariant, spin conserving interaction $V(\bm{r},\bm{r'})$ given by
\begin{align}
    \label{matrix element bloch states expanded 3d}
    V&(\bk,\bk',\bq,\sigma,\sigma')\\
    =&\bra{\Qm(\bk+\bq),\sigma,\Qm(\bk'-\bq),\sigma'}\hat{V}\ket{\bk,\sigma,\bk',\sigma'}\nl
    =&\frac{1}{N^2}\iint_{V}d\bm{r}d\bm{r}'
    \bm{u}^\dagger_{\sigma,\bk+\bm{q}}(\bm{r})\bm{u}_{\sigma,\bk}(\bm{r})e^{-i\bq\cdot\bm{r}}\nl
    &\bm{u}^\dagger_{\sigma',\bk'-\bm{q}}(\bm{r}')\bm{u}_{\sigma',\bk'}(\bm{r}')e^{i\bq\cdot\bm{r}'}V(\hat{\bm{r}},\hat{\bm{r}}')\,.\nonumber
\end{align}
Inserting the Fourier expansion of \cref{main-screened interaction real space} and using the definition of the overlap integrals of \cref{main-overlap factors} results in
\begin{align}
    V&(\bk,\bk',\bq,\sigma,\sigma')\nl
    =&\frac{1}{N\Omega L}\sum_{\bq',\bG,\bG'}V_{\bG,\bG'}(\bq')\nl
    &\bigg(\frac{1}{N}\sum_{\bm{R}}e^{i(\bq'-\bq)\cdot\bm{R}}\bigg) \bigg(\frac{1}{N}\sum_{\bm{R}'}e^{i(\bq-\bq')\cdot\bm{R}'}\bigg)\nl
    &\Fm^{\sigma}_{\bk+\bq,\bk}(\bq-\bq'-\bG) \Fm^{\sigma'}_{\bk'-\bq,\bk'}(\bq'-\bq+\bG')\nl
    =&\frac{1}{\mathcal{V}}\sum_{\bG,\bG'}\sum_{q_z}V_{\bm{G},\bm{G}'}(\bq+q_z\hat{z};\omega)\nl
    &\Fm^{\sigma}_{\bk+\bq,\bk}(-q_z'\hat{z}-\bm{G}) \Fm^{\sigma'}_{\bk'-\bq,\bk'}(q_z'\hat{z}+\bm{G}')\,.
    \label{matrix element general coulomb interaction}
\end{align}
The sums in the brackets ensure $\bq'_{\parallel}=\bq$ but leave $q'_z$ undetermined.
Assuming that the orbitals are tightly bound to the plane, we have $r_z\approx0$ in \cref{main-overlap factors} where the integral has significant support;  as such we can approximate $\Fm(\bG+q_z'\hat{z})\approx\Fm(\bG)$.
Performing the sum over $q_z'$ on the Fourier coefficients yields the 2D interaction 
\begin{align}
\label{two dimensional screened interaction tensor}
   V_{\bm{G},\bm{G}'}^{\T{2D}}(\bq;\omega)=\frac{1}{L}\sum_{q_z}V_{\bm{G},\bm{G}'}(\bq+q_z\hat{z};\omega)\,.
\end{align}
Applying the above steps to the screened Coulomb interaction yields \cref{main-matrix element bloch states expanded}.
For the unscreened interaction $V_0$, we use translational invariance to find that the expansion coefficients have the form 
\begin{align}
    (V_0)_{\bG,\bG'}(\bq)=\delta_{\bG,\bG'}V_0(\bq+\bG),
\end{align}
with $V_0(\bq)$ the Fourier transform of the three-dimensional unscreened Coulomb potential relative to the origin.
Inserting this into \cref{matrix element general coulomb interaction} and taking $L\rightarrow\infty$, we can convert the sum over $q'_z$ into an integral and find
\begin{align}
    \label{coulomb interaction bloch states 2d}
    V_0&(\bk,\bk',\bq,\sigma,\sigma')\\
    =&\frac{1}{N\Omega L}\sum_{\bm{G}} \sum_{q'_z}V_{0}(\bm{q}+\bm{G}+q'_z\bm{z})\nl 
    &\Fm^{\sigma}_{\bk+\bq,\bk}(-\bm{G}) \Fm^{\sigma'}_{\bk'-\bq,\bk'}(\bm{G})\nl
    =&\frac{1}{N\Omega}\sum_{\bm{G}}\frac{1}{2\pi}\int_{-\infty}^{\infty}dq'_z V_{0}(\bm{q}+\bm{G}+q'_z\bm{z})\nl 
    &\Fm^{\sigma}_{\bk+\bq,\bk}(-\bm{G}) \Fm^{\sigma'}_{\bk'-\bq,\bk'}(\bm{G})\nl
    =&\frac{1}{N\Omega}\sum_{\bm{G}}V^{\T{2D}}_0(\bm{q}+\bm{G})\Fm^{\sigma}_{\bk+\bq,\bk}(-\bm{G}) \Fm^{\sigma'}_{\bk'-\bq,\bk'}(\bm{G})\,,\nonumber
\end{align}
where $V^{\T{2D}}_0(\bq)$ is the Fourier transform of the unscreened Coulomb interaction in two-dimensions as introduced in the main text.
In the same continuum limit $L\rightarrow\infty$ and using again the approximation of tightly bound charges in the Fourier transform of the Dyson equation \cref{main-Dyson equation wrpa}, we find
\begin{align}
    \label{dyson equation 2d rpa}
    &V_{\bm{G},\bm{G}'}^{\T{2D,RPA}}(\bq,\omega)=\delta_{\bG,\bG'}V^{\T{2D}}_0(\bq+\bG)\nl
    &+V^{\T{2D}}_0(\bq+\bG)\sum_{\bG''}P_{\bG,\bG''}(\bq;\omega)V_{\bm{G}'',\bm{G}'}^{\T{2D,RPA}}(\bq,\omega)\,,
\end{align}
which is the two-dimensional analog of the usual RPA Dyson equation in the presence of a lattice as reported in the literature~\cite{hybertsenElectronCorrelationSemiconductors1986}.
The elements of the polarizability matrix entering \cref{dyson equation 2d rpa} are given by \cref{main-susceptibility}.
With the above, we recast the RPA screening for a two-dimensional material into a purely two-dimensional problem while retaining the three-dimensional nature of the overlap integrals $\Fm$.
\cref{dyson equation 2d rpa} is a matrix equation in the space of the reciprocal lattice vectors $\{\bG\}$ whose formal solution is given by \cref{main-formal solution dyson equation in 2d}.
\nolinenumbers
\begin{figure*}[ht]
    \centering
    \subfloat
    {
        \include{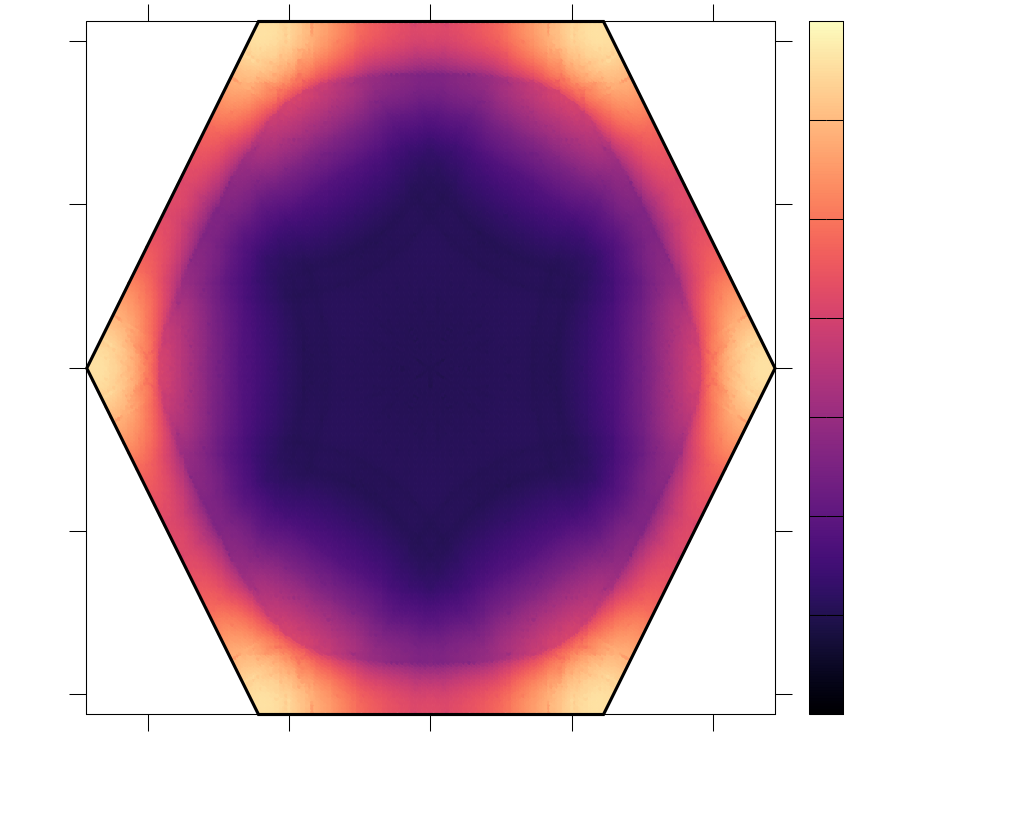}
    }
    \subfloat
    {
        \include{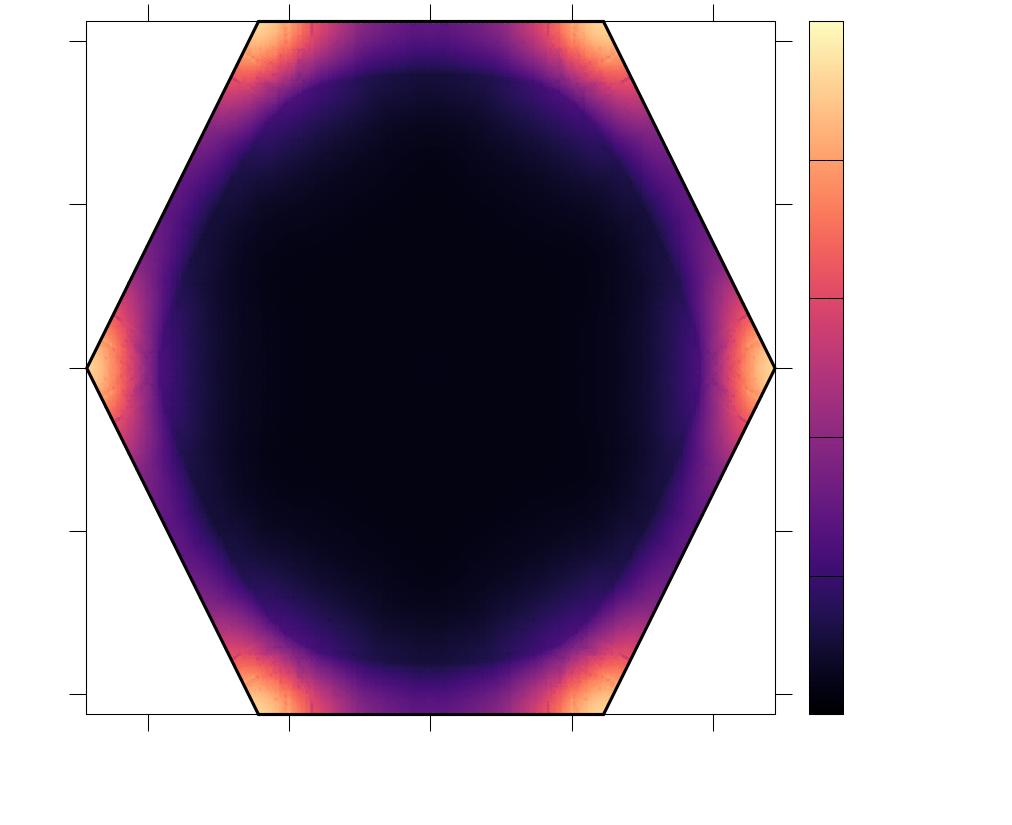}
    }
    \caption{
    Macroscopically screened interaction potential without Umklapp corrections $V^{\T{RPA}}_{\T{macro}}$  (a) and enhancement upon including corrections due to Umklapp scattering by the reciprocal lattice up to the third nearest neighbor of the origin (b).
    The backfolding of adjacent BZs results in an enhancement of the interaction for large $\bm{q}$ as opposed to small $\bm{q}$.
    }
    \label{fig: screened interaction with Umklapp}
\end{figure*}
\subsection{Diagrammatic derivation of the RPA screening \label{app: rpa feynman}}

An alternative derivation of the screened interaction $\hat{V}_{\T{RPA}}$ on a two-dimensional lattice is given by a diagrammatic expansion in the Matsubara framework~\cite{bruusManyBodyQuantumTheory2004}. 
We define the interaction vertex for the unscreened interaction
\begin{align}
&
    \begin{tikzpicture}[node distance = 1.25cm]
    \def\labeldist{0.1cm};
        \begin{feynhand}
            \vertex [dot] (v0) at (-1,0){};
            \vertex [dot] (v1) at (1,0){};
            \propag [photon] (v0) to [edge label = ${q,iq_{n}}$ ] (v1);
            \vertex [above right = \labeldist of v1,align=left] {$\Qm(\bk'-\bq),$\\$ik'_n-iq_n,\sigma'$};
            \vertex [below right = \labeldist of v1] {$\bk',ik'_n,\sigma'$};
            \vertex [above left = \labeldist of v0,align=left] {$\Qm(\bk+\bq),$\\$ik_n+iq_n,\sigma$};
            \vertex [below left = \labeldist of v0] {$\bk,ik_n,\sigma$};
            \vertex [below left = 0.2cm of v0] (in1) {};
            \vertex [above left = 0.2cm of v0] (out1) {};
            \vertex [below right = 0.2cm of v1] (in2) {};
            \vertex [above right = 0.2cm of v1] (out2) {};
            \draw  (v0) -- (in1);
            \draw  (v0) -- (out1);
            \draw  (v1) -- (in2);
            \draw  (v1) -- (out2);
        \end{feynhand}
    \end{tikzpicture}\nl
    &=-N\Omega V_0^{\T{2D}}(\bk,\bk',\bq,\sigma,\sigma')\,.
    \label{diagram free interaction}
\end{align}
The second ingredient necessary for writing the diagrammatic expansion of the screened interaction is the free propagator for Bloch electrons
\begin{equation}
    \begin{tikzpicture}
    \def\labeldist{0.1cm};
        \begin{feynhand}
            \vertex (v0) at (0,0);
            \vertex (v1) at (3,0){$=G_0(\bk,ik_n,\sigma)=\displaystyle\frac{1}{ik_n-\xi_{\bk,\sigma}}$};
            \propag [fer] (v0) to [edge label = ${\bk,ik_{n},\sigma}$ ] (v1);
        \end{feynhand}
    \end{tikzpicture}.
\end{equation}
The full Dyson equation for the screened interaction is a function of the irreducible pair bubble~\cite{bruusManyBodyQuantumTheory2004} as
\begin{equation}
\begin{tikzpicture}[ node distance=1cm]
    \begin{feynhand}
        \vertex [dot] (v0) at (0,0) {};
        \vertex [dot,right = 1cm of v0] (v1) {};
        \draw [double, decorate, decoration={coil,aspect= 0}] (v0) to (v1);
        \vertex [below left = 0.2cm of v0] (in0) {};
        \vertex [above left = 0.2cm of v0] (out0) {};
        \vertex [below right = 0.2cm of v1] (in1) {};
        \vertex [above right = 0.2cm of v1] (out1) {};
        \draw  (v0) -- (in0);
        \draw  (v0) -- (out0);
        \draw  (v1) -- (in1);
        \draw  (v1) -- (out1);
        \vertex [right = 0.05cm of v1] (v20) {$=$};
        \vertex [dot, right = 0.05cm of v20] (v00) {};
        \vertex [dot,right = 0.75cm of v00] (v10) {};
        \propag[photon] (v00) to (v10);
        \vertex [below left = 0.2cm of v00] (in00) {};
        \vertex [above left = 0.2cm of v00] (out00) {};
        \vertex [below right = 0.2cm of v10] (in10) {};
        \vertex [above right = 0.2cm of v10] (out10) {};
        \draw  (v00) -- (in00);
        \draw  (v00) -- (out00);
        \draw  (v10) -- (in10);
        \draw  (v10) -- (out10);
        \vertex [right = 0.05cm of v10] (v2) {$+$};
        \vertex [dot,right = 0.55cm of v2.west] (midv2) {};
        \vertex [below right = 0.2cm of v2.west] (in2) {};
        \vertex [above right = 0.2cm of v2.west] (out2) {};
        \draw  (midv2) -- (in2);
        \draw  (midv2) -- (out2);
        \vertex[ right = -0.15cm of v2.east, anchor=west] (v6){};
        \vertex[dot,right = 0.5cm of v6](v7){};
        \propag[photon] (v6) to (v7);
        \vertex[dot,right = 0.75cm of v7.center, anchor=center] (v8){};
        \propag[fer,in=275,out=265,looseness=1.3] (v8) to (v7);
        \propag[fer,in=95,out=85,looseness=1.3] (v7) to (v8);
        \vertex[ right=0.375cm of v7.center, anchor=center] (vb){};
        \vertex[NEblob] at (vb){};
        \vertex [right = -0.25cm of v8.east, anchor=west] (v3) {};
        \vertex [dot, right = 1cm of v3] (v4) {};
        \draw [double, decorate, decoration={coil,aspect= 0}] (v3) to (v4);
        \vertex [below right = 0.2cm of v4] (in4) {};
        \vertex [above right = 0.2cm of v4] (out4) {};
        \draw  (v4) -- (in4);
        \draw  (v4) -- (out4);
    \end{feynhand}
\end{tikzpicture}.
\label{dyson equation full}
\end{equation}
The RPA is equivalent to approximating the irreducible pair bubble with the lowest order diagram contributing to it.
The resulting pair bubble is the free polarizability and the associated Dyson equation has the formal solution
\begin{equation}
\begin{tikzpicture}[ node distance=1cm]
    \begin{feynhand}
        \vertex [dot] (v0) at (0,0) {};
        \vertex [dot,right = 1cm of v0] (v1) {};
        \draw [double, decorate, decoration={coil,aspect= 0}] (v0) to [edge label = $\T{RPA}$] (v1);
        \vertex [below left = 0.2cm of v0] (in0) {};
        \vertex [above left = 0.2cm of v0] (out0) {};
        \vertex [below right = 0.2cm of v1] (in1) {};
        \vertex [above right = 0.2cm of v1] (out1) {};
        \draw  (v0) -- (in0);
        \draw  (v0) -- (out0);
        \draw  (v1) -- (in1);
        \draw  (v1) -- (out1);
        \vertex [right = 0.05cm of v1] (v2) {$=$};
        \vertex [right = 0.05cm of v2] (v2p) {$\bigg(1-$};
        \vertex [dot,right = 0.55cm of v2.west] (midv2) {};
        \vertex [below right = 0.2cm of v2.west] (in2) {};
        \vertex [above right = 0.2cm of v2.west] (out2) {};
        \draw  (midv2) -- (in2);
        \draw  (midv2) -- (out2);
        \vertex[ right = -0.15cm of v2p.east, anchor=west] (v6){};
        \vertex[dot,right = 0.5cm of v6](v7){};
        \propag[photon] (v6) to (v7);
        \vertex[dot,right = 0.75cm of v7.center, anchor=center] (v8){};
        \propag[fer,in=315,out=225,looseness=1] (v8) to (v7);
        \propag[fer,in=135,out=45,looseness=1] (v7) to (v8);
        \node[right=0.05 cm of v8.east, anchor=west](v10){$\bigg)^{-1}$};
        \vertex [right = -0.25cm of v10.east, anchor=west] (v3) {};
        \vertex [dot, right = 0.75cm of v3] (v4) {};
        \propag[photon] (v3) to (v4);
        \vertex [below right = 0.2cm of v4] (in4) {};
        \vertex [above right = 0.2cm of v4] (out4) {};
        \draw  (v4) -- (in4);
        \draw  (v4) -- (out4);
        \vertex[below right= -0.05cm and 0.05cm of v4] {\,,};
    \end{feynhand}
\end{tikzpicture}
\label{dyson rpa diagram}
\end{equation}
from which the matrix element of the static screened interaction can be read out as
\begin{align}
    &\begin{tikzpicture}[node distance = 1.25cm]
    \def\labeldist{0.1cm};
        \begin{feynhand}
            \vertex [dot] (v0) at (-1,0){};
            \vertex [dot] (v1) at (1,0){};
            \draw [double, decorate, decoration={coil,aspect= 0}] (v0) to [edge label = ${\T{RPA},\bq,i0^+}$] (v1);
            \vertex [above right = \labeldist of v1,align=left] {$\Qm(\bk'-\bq),i0^+,\sigma'$};
            \vertex [below right = \labeldist of v1] {$\bk',i0^+,\sigma'$};
            \vertex [above left = \labeldist of v0,align=left] {$\Qm(\bk+\bq),i0^+,\sigma$};
            \vertex [below left = \labeldist of v0] {$\bk,i0^+,\sigma$};
            \vertex [below left = 0.2cm of v0] (in1) {};
            \vertex [above left = 0.2cm of v0] (out1) {};
            \vertex [below right = 0.2cm of v1] (in2) {};
            \vertex [above right = 0.2cm of v1] (out2) {};
            \draw  (v0) -- (in1);
            \draw  (v0) -- (out1);
            \draw  (v1) -- (in2);
            \draw  (v1) -- (out2);
        \end{feynhand}
    \end{tikzpicture}\nl
    &=-N\Omega V(\bk,\bk',\bq,\sigma,\sigma').
    \label{diagram definition rpa}
\end{align}

The inverse of the term in brackets in \cref{dyson rpa diagram} is also to be taken in the degrees of freedom of the reciprocal lattice, $\bm{G},\bm{G}'$.
This can be made explicit by introducing an unscreened interaction vertex of matrix form in these indices as
\begin{align}
    &
    \begin{tikzpicture}[node distance = 1.25cm]
    \def\labeldist{0.1cm};
        \begin{feynhand}
            \vertex [dot] (v0) at (-1,0){};
            \vertex [above =\labeldist of v0] {$\bm{G}$};
            \vertex [dot] (v1) at (1,0){};
            \propag [scalar] (v0) to [edge label = ${q,iq_{n}}$ ] (v1);
            \vertex [above right = \labeldist of v1,align=left] {$\Qm(\bk'-\bq),$\\$ik'_n-iq_n,\sigma'$};
            \vertex [below right = \labeldist of v1] {$\bk',ik'_n,\sigma'$};
            \vertex [above left = \labeldist of v0,align=left] {$\Qm(\bk+\bq),$\\$ik_n+iq_n,\sigma$};
            \vertex [below left = \labeldist of v0] {$\bk,ik_n,\sigma$};
            \vertex [above =\labeldist of v1] {$\bm{G}'$};
            \vertex [below left = 0.2cm of v0] (in0) {};
            \vertex [above left = 0.2cm of v0] (out0) {};
            \vertex [below right = 0.2cm of v1] (in1) {};
            \vertex [above right = 0.2cm of v1] (out1) {};
            \draw  (v0) -- (in0);
            \draw  (v0) -- (out0);
            \draw  (v1) -- (in1);
            \draw  (v1) -- (out1);
        \end{feynhand}
    \end{tikzpicture}\nl
    &=-V_0^{\T{2D}}(\bm{q}+\bm{G})\Fm^{\sigma}_{\bk+\bq,\bk}(-\bm{G}) \Fm^{\sigma '}_{\bk'-\bq,\bk'}(\bm{G}')\,.
    \label{diagram interaction separated}
\end{align}
The unscreened interaction vertex proper is the trace of this matrix as
\begin{align}
    \begin{tikzpicture}[node distance = 1.cm]
    \def\labeldist{0.1cm};
        \begin{feynhand}
            \vertex [dot] (v0) at (0,0){};
            \vertex [dot, right = of v0] (v1){};
            \propag [photon] (v0) to [edge label = ${q,iq_{n}}$ ] (v1);
            \vertex [right = 0.cm of v1] (v2) {$=$};
            \vertex [dot, right = of v2] (v3) {};
            \vertex [dot, right = of v3] (v4) {};
            \propag [scalar] (v3) to [edge label = ${q,iq_{n}}$ ](v4);
            \vertex [left = 0.cm of v3] {$\T{Tr}_{\bm{G}}\lbrace$};
            \vertex [right = 0.cm of v4] {$\rbrace\,.$};
            \vertex [below left = 0.2cm of v0] (in0) {};
            \vertex [above left = 0.2cm of v0] (out0) {};
            \vertex [below right = 0.2cm of v1] (in1) {};
            \vertex [above right = 0.2cm of v1] (out1) {};
            \draw  (v0) -- (in0);
            \draw  (v0) -- (out0);
            \draw  (v1) -- (in1);
            \draw  (v1) -- (out1);
            \vertex [below left = 0.2cm of v3] (in3) {};
            \vertex [above left = 0.2cm of v3] (out3) {};
            \vertex [below right = 0.2cm of v4] (in4) {};
            \vertex [above right = 0.2cm of v4] (out4) {};
            \draw  (v3) -- (in3);
            \draw  (v3) -- (out3);
            \draw  (v4) -- (in4);
            \draw  (v4) -- (out4);
        \end{feynhand}
    \end{tikzpicture}
\end{align}

The usefulness of \cref{diagram interaction separated} becomes apparent upon realizing that at any order in the Dyson equation, we may rearrange the vertices as 
\begin{equation}
\label{exchanging vertices}
\begin{tikzpicture}[ node distance=1.25cm]
    \begin{feynhand}
        \vertex [dot] (v0) at (0,0cm) {};
        \vertex[dot,right = 0.75 cm of v0](v1){};
        \propag[photon] (v0) to [edge label = ${\bm{q},iq_n}$] (v1);
        \vertex[dot,right = 0.5cm of v1.east, anchor=west] (v2){};
        \vertex[dot,right= 0.75cm of v2](v3){};
        \propag[photon] (v2) to [edge label = ${\bm{q},iq_n}$] (v3);
        \propag[fer,in=315,out=225] (v2) to (v1);
        \propag[fer,in=135,out=45] (v1) to (v2);
        \vertex[right = 0.25cm of v3.east, anchor=west](v4){$=\sum_{\bm{G},\bm{G}'}$};
        \vertex [dot,right = 0.0cm of v4.east, anchor=west ] (v5) {};
        \vertex[below = 0.00 of v5 ]{$\bm{G}$};
        \vertex[dot,right = 0.75cm of v5](v6){};
        \vertex[below = 0.00 of v6 ]{$\bm{G}$};
        \propag[scalar] (v5) to [edge label = ${\bm{q},iq_n}$] (v6);
        \vertex[dot,right = 0.5cm of v6.east, anchor=west] (v7){};
        \vertex[below = 0.00 of v7 ]{$\bm{G}'$};
        \vertex[dot,right= 0.75cm of v7](v8){};
        \vertex[below = 0.00 of v8 ]{$\bm{G}'$};
        \propag[scalar] (v7) to [edge label = ${\bm{q},iq_n}$] (v8);
        \propag[fer,in=315,out=225] (v7) to (v6);
        \propag[fer,in=135,out=45] (v6) to (v7);
        \vertex[below left = 1.25cm and 0.3cm of v0.center, anchor=west](v9){$=\sum_{\bm{G},\bm{G}'}$};
        \vertex[dot, right= -0.05cm of v9] (v10){};
        \vertex[above left = -0.1cm and -0.2cm of v10 ]{$\bm{G}'$};
        \vertex[dot, right = 0.8cm of v10] (v11){};
        \vertex[above right = -0.1cm and -0.15cm of v11 ]{$\bm{G}$};
        \propag[fer,out=225,in=315,looseness=1] (v11) to (v10);
        \propag[fer,out=90,in=90,looseness=1.5] (v10) to (v11);
        \propag[scalar] (v10) to [edge label= ${\bm{q},iq_n}$ ] (v11);
        \vertex[dot,right=0.45cm of v11](v12){};
        \vertex[below = 0.00 of v12 ]{$\bm{G}$};
        \vertex[dot, right = 0.75cm of v12] (v13){};
        \vertex[below = 0.00 of v13 ]{$\bm{G}'$};
        \propag[scalar] (v12) to [edge label = ${\bm{q},iq_n}$] (v13);
        \vertex[right = 0.05cm of v13.east, anchor= west](v14){$=\T{Tr}_{\bm{G}}\bigg\lbrace$};
        \vertex[dot, right= -0.15cm of v14] (v15){};
        \vertex[dot, right = 0.8cm of v15] (v16){};
        \propag[fer,out=225,in=315,looseness=1] (v16) to (v15);
        \propag[fer,out=90,in=90,looseness=1.5] (v15) to (v16);
        \propag[scalar] (v15) to [edge label= ${\bm{q},iq_n}$ ] (v16);
        \vertex[dot,right=0.2cm of v16](v17){};
        \vertex[dot, right = 0.75cm of v17] (v18){};
        \propag[scalar] (v17) to [edge label = ${\bm{q},iq_n}$] (v18);
        \vertex[right=0.cm of v18]{$\bigg\rbrace\,.$};
        \vertex [below left = 0.2cm of v0] (in0) {};
        \vertex [above left = 0.2cm of v0] (out0) {};
        \vertex [below right = 0.2cm of v3] (in3) {};
        \vertex [above right = 0.2cm of v3] (out3) {};
        \draw  (v0) -- (in0);
        \draw  (v0) -- (out0);
        \draw  (v3) -- (in3);
        \draw  (v3) -- (out3);
        \vertex [below left = 0.2cm of v5] (in5) {};
        \vertex [above left = 0.2cm of v5] (out5) {};
        \vertex [below right = 0.2cm of v8] (in8) {};
        \vertex [above right = 0.2cm of v8] (out8) {};
        \draw  (v5) -- (in5);
        \draw  (v5) -- (out5);
        \draw  (v8) -- (in8);
        \draw  (v8) -- (out8);
        \vertex [below left = 0.2cm of v12] (in12) {};
        \vertex [above left = 0.2cm of v12] (out12) {};
        \vertex [below right = 0.2cm of v13] (in13) {};
        \vertex [above right = 0.2cm of v13] (out13) {};
        \draw  (v12) -- (in12);
        \draw  (v12) -- (out12);
        \draw  (v13) -- (in13);
        \draw  (v13) -- (out13);
        \vertex [below left = 0.2cm of v17] (in17) {};
        \vertex [above left = 0.2cm of v17] (out17) {};
        \vertex [below right = 0.2cm of v18] (in18) {};
        \vertex [above right = 0.2cm of v18] (out18) {};
        \draw  (v17) -- (in17);
        \draw  (v17) -- (out17);
        \draw  (v18) -- (in18);
        \draw  (v18) -- (out18);
    \end{feynhand}
\end{tikzpicture}
\end{equation}
Summing up the internal Matsubara frequencies in the interacting pair bubble above yields the same product of the unscreened interaction with an element of the polarizability tensor as found in \cref{main-dielectric tensor}~\cite{bruusManyBodyQuantumTheory2004}.
Solving the Dyson equation with the above rewriting recasts \cref{diagram definition rpa} into the form
\begin{equation}
\begin{tikzpicture}[ node distance=1cm]
    \begin{feynhand}
        \vertex [dot] (v0) at (0,0) {};
        \vertex [dot,right = of v0] (v1) {};
        \draw [double, decorate, decoration={coil,aspect= 0}] (v0) to [edge label = $\T{RPA}$] (v1);
        \vertex [right = 0.25cm of v1] (v2) {$=\T{Tr}_{\bm{G}}\bigg\lbrace\bigg(1-$};
        \vertex[dot, right= 0.cm of v2] (v3){};
        \vertex[dot, right = 0.5cm of v3] (v4){};
        \propag[fer,out=225,in=315,looseness=1] (v4) to (v3);
        \propag[fer,out=45,in=135,looseness=1] (v3) to (v4);
        \propag[scalar] (v3) to (v4); 
        \node[right=0.cm of v4](v5){$\bigg)^{-1}$};
        \vertex [dot, right = -0.25cm of v5] (v6) {};
        \vertex [dot, right = of v6] (v7) {};
        \propag[scalar] (v6) to (v7);
        \vertex[right=0.cm of v7]{$\bigg\rbrace\,,$};
        \vertex [below left = 0.2cm of v0] (in0) {};
        \vertex [above left = 0.2cm of v0] (out0) {};
        \vertex [below right = 0.2cm of v1] (in1) {};
        \vertex [above right = 0.2cm of v1] (out1) {};
        \draw  (v0) -- (in0);
        \draw  (v0) -- (out0);
        \draw  (v1) -- (in1);
        \draw  (v1) -- (out1);
        \vertex [below left = 0.2cm of v6] (in6) {};
        \vertex [above left = 0.2cm of v6] (out6) {};
        \vertex [below right = 0.2cm of v7] (in7) {};
        \vertex [above right = 0.2cm of v7] (out7) {};
        \draw  (v6) -- (in6);
        \draw  (v6) -- (out6);
        \draw  (v7) -- (in7);
        \draw  (v7) -- (out7);
    \end{feynhand}
\end{tikzpicture}
\label{rpa simplfied diagrammatics}
\end{equation}
which is the diagrammatic representation of \cref{main-matrix element bloch states expanded,main-formal solution dyson equation in 2d}.

\subsection{Screening from normal processes}
 
The dependence on the inverse dielectric tensor showcases that in principle {\it all} elements of the dielectric tensor, and conversely all entries of the polarizability tensor $P_{\bG,\bG'}$, are required for the full calculation of the screened interaction between Bloch states.
However, the finite $\bG$ entries of \cref{main-matrix element bloch states expanded} correspond to Umklapp scattering involving the momentum exchange of at least one reciprocal lattice vector.
Their importance can be estimated by the impact of the lattice spacing on the overlap integrals.
For lattice spacings $a$ which are small compared to the extent of the considered orbitals, the effect of the lattice vanishes and one recovers an approximate system of free electrons like, e.g., in the simple metals~\cite{sinhaMicroscopicTheoryDielectric1974}.
In this scenario, the product of spinors in \cref{main-overlap factors} is nearly homogeneous within the unit cell, resulting in a suppression of all $\Fm(\bG)$ with finite $\bG$.
Compare that to the limit of very large $a$, where the orbitals are strongly localized in the unit cell, resulting in $\Fm(\bG)\approx \Fm(0)$ for large swathes of the reciprocal lattice around the origin $\bG=\bm{0}$.
While inevitably $\Fm(\bG)\rightarrow0$ as $||\bG||\rightarrow\infty$, Umklapp scattering plays a dominant role in this case.
Monolayer NbSe$_2$ lies in an intermediate regime between these two extreme cases.
The interlaced Se sublattices increase the spacing of the Nb sublattice, such that Umklapp scattering is quite prominent for a significant number of reciprocal lattice momenta surrounding the origin of the reciprocal lattice.
To visualize the impact this has, we compare against a first approximation in which one drops the Umklapp processes in \cref{main-matrix element bloch states expanded} and restricts to normal processes ($\bG=\bm{0}$).
This case of considering only normal processes is related to introducing a macroscopic dielectric function~\cite{dolgovAdmissibleSignStatic1981} and is not justified for ML NbSe$_2$, with significant corrections by Umklapp processes present in this material.
The error incurred by this approximation as well as models to extrapolate between it and the proper tensorial treatment of the polarizability have been discussed in the literature~\cite{sinhaMicroscopicTheoryDielectric1974,dolgovAdmissibleSignStatic1981,johannesFermisurfaceNestingOrigin2006}.

Calculating \cref{main-susceptibility} for $\bG=\bG'=\bm{0}$ reproduces the expression for the polarizability of a real solid in the limit of neglecting local field effects~\cite{ehrenreichSelfConsistentFieldApproach1959,cookeElectronicSusceptibilityNiobium1974,johannesFermisurfaceNestingOrigin2006,kimQuasiparticleEnergyBands2017}.
Inserting this macroscopic static polarizability $P_{\bm{0},\bm{0}}(\bq;0)$ into \cref{main-matrix element bloch states expanded} yields 
\begin{align}
    \label{w rpa truncated to 0}
    &V(\bk,\bk',\bq,\sigma,\sigma')=\frac{1}{N\Omega}V^{\T{RPA}}_{\T{macro}}(\bm{q})\Fm^{\sigma}_{\bk+\bq,\bk}(\bm{0}) \Fm^{\sigma'}_{\bk'-\bq,\bk'}(\bm{0})\,,
\end{align}
where we identify the macroscopically screened interaction potential \footnote{
    Approximating the $\bm{q}$-dependent polarizability in \cref{screened interaction potential} with its value at the $\bm{\Gamma}$ point reproduces the Thomas-Fermi screening formula for two dimensions.
}
\begin{align}
   \label{screened interaction potential}
   V^{\T{RPA}}_{\T{macro}}(\bm{q})&=\frac{V^{2D}_0(\bm{q})}{1-V^{\T{2D}}_0(\bm{q})P_{\bm{0},\bm{0}}(\bm{q})}\\
   &=\frac{e^2}{2\epsilon_0}\bigg(|\bm{q}|-\frac{e^2}{2\epsilon_0}P_{\bm{0},\bm{0}}(\bm{q})\bigg)^{-1}\,.\nonumber
\end{align}

The macroscopic noninteracting polarizability is shown in \cref{main-fig:screened} (a) for the parametrization introduced in \cref{tight binding model}.
As discussed, e.g., by \citeauthor{johannesFermisurfaceNestingOrigin2006}~\cite{johannesFermisurfaceNestingOrigin2006}, the lower-lying bands not included in the tight-binding model we employ should give a small quantitative correction to the polarizability calculated here.
\cref{fig: screened interaction with Umklapp}(a) shows the $\bm{q}$-dependent macroscopically screened interaction potential $V^{\T{RPA}}_{\T{macro}}$.
\nolinenumbers
\begin{figure*}[ht]
    \centering
    \subfloat[]
    {
        \includegraphics[width=0.9\columnwidth]{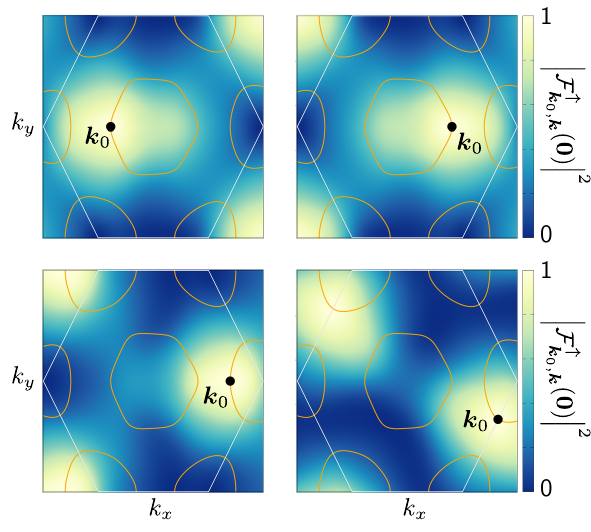}
    }
    \subfloat[]
    {
        \includegraphics[width=0.9\columnwidth]{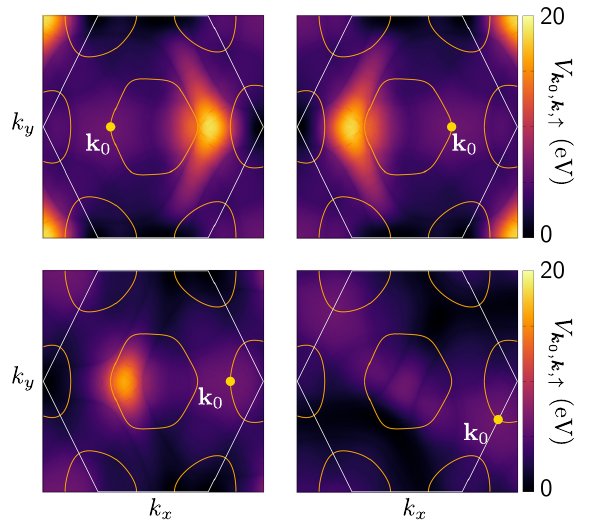}
    }
    \caption{
    Shown are the overlap $|\Fm^{\up}_{\bm{k},\bm{k}_0}(\bm{0})|^2$ (a) and the matrix elements $V_{\bk,\bk_0,\up}$ (b) for selected initial state $\bm{k}_0$ on the $\Gamma$ and $K$ Fermi surfaces as a function of the target state $\bk$. The stronger interaction for large-momentum transfers, where the overlaps in general diminish, results in a strong selectivity of the matrix elements throughout the range of available target states.
    }
    \label{fig: form factors general}
\end{figure*}
\subsection{Screening including Umklapp processes\label{subsec: umklapp}}

The screening from normal processes is the dominant contribution for simple metals.
Numerically evaluating the overlaps $\Fm(\bG)$ for $\bG=n\bm{b}_1+m\bm{b}_2$ up to $n,m>40$ shows that they only drop below 10\% of their maximum value at $|n|>17$ along $\bm{b}_1$, and similarly for $\bm{b}_2$.
Luckily, numerical convergence for calculating the screened interaction can be reached by considering far fewer reciprocal lattice sites, as the corresponding unscreened interaction $V^{\T{2D}}_{0}(\bq+\bG)$ vanishes as $1/||\bG||$ for larger $\bG$.
We find that at least the first few nearest neighbor sites from the origin have non-negligible impact on the result.
This is due to the fact that close to the Bragg planes delimiting the first BZ we have by definition ${V_{\T{2D}}(\bm{q}-\bm{G})\approx V_{\T{2D}}(\bm{q})}$ for one nearest neighbor $\bG$.
In this region around the boundary of the first BZ no truncation due to either the smallness of the unscreened interaction or a vanishing overlap is justified.
Hence, we numerically solve \cref{main-matrix element bloch states expanded} including up to the third nearest neighbors.
Note that the contributions from further reciprocal lattice sites increasingly become $\bq$-independent, as both the $\bq$-dependent features of the polarizability and the bare interaction are washed out due to $V_{\T{2D}}(\bm{q}-\bm{G})\approx e^2/(2\epsilon_0 |G|)$.
As such, their small quantitative contribution will not contribute to the mechanism for superconductivity that we discuss below.

Due to the overlaps occurring in \cref{main-matrix element bloch states expanded}, the screened interaction with Umklapp scattering included can no longer be split into a $\bq$-dependent interaction potential and an overlap term depending on the involved states.
However, for ML NbSe$_2$ the overlaps $\Fm(\bG)$ for the first few nearest neighbor (NN) reciprocal lattice sites are well approximated by $\Fm(\bG_{\T{NN}})\approx\Fm(\bm{0})$.
Using the effective interaction of \cref{main-matrix element simplified} with a truncation to the third nearest neighbors of the first BZ~\footnote{
Comparison to a direct numerical evaluation of the matrix elements indicates an error $<10\%$ in this approximation.
}
, we are again able to identify a $\bm{q}$-dependent effective screened interaction potential $V^{\T{RPA}}(\bq)$ as in the case without Umklapp scattering.
The main effect of the Umklapp scattering in this approximation becomes the folding back of the interaction potential from adjacent BZs onto the first BZ. 
This folding back selectively enhances the interaction for large momenta closer to the boundary of the first BZ.
The resulting enhancement of the interaction potential inside the first BZ is shown in \cref{fig: screened interaction with Umklapp}(b).

\subsection{Pair scattering for Kramers pairs}

The relevant matrix elements we consider are those describing the scattering of Kramers partners given by
\begin{align}
    \label{V Kramers pair}
    &V_{\bk,\bk',\up}\approx \frac{1}{N\Omega}V^{\T{RPA}}(\bk-\bk')\left|\Fm^{\up}_{\bk,\bk'}(\bm{0})\right|^2\,. 
\end{align}
The product of the overlaps reduces to a modulus square because the spinors belonging to time-reversed states obey $\bm{u}_{\bar{\sigma},\bar{\bk}} = \bm{u}_{\sigma,{\bk}}^*$.
The overlap $|\Fm_{\bm{k},\bm{k}_0}^\uparrow(\bm{0})|^2$ between Bloch spinors at the points $\bm{k}$ and $\bm{k}_0$ is shown for selected $\bm{k}_0$ in the three Fermi pockets in \cref{fig: form factors general}(a).

\begin{figure}[ht]
    \begin{center}
    \includegraphics[width=\columnwidth]{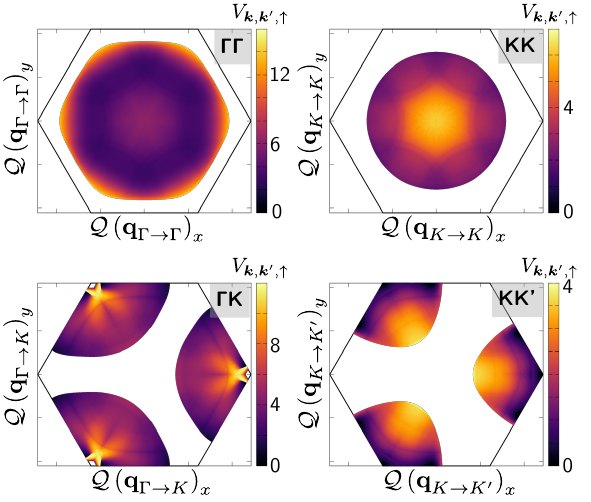}
    \end{center}
    \caption{\label{fig: matrix elements FStoFS}
    Possible momentum transfers $\Qm(\bm{q})$ within, and between, the $\Gamma$ and $K$ Fermi surfaces. The value of the matrix elements $V_\mathrm{KP}(\bm{k},\bm{k}')$ is shown in color scale.
    }
\end{figure}

Both the overlap and the interaction potential for a specific momentum exchange enter into the final matrix element.
Therefore, the Coulomb interaction becomes selective, scattering an electron in a given initial state only into a few regions within the Brillouin zone with the largest matrix elements.
As an example, we show in \cref{fig: form factors general}(b) the highly anisotropic matrix elements throughout the entire BZ.
This anisotropy plays a major role for the pairing mechanism discussed below as it favors multi patch superconductivity.
At the same time, the strong selectivity throughout the BZ will contribute to the importance of nematic and chiral solutions that are less prominent for more isotropic interactions~\cite{wanObservationSuperconductingCollective2022}.

To gain some insight into the allowed gap symmetries and the relationships between gaps at different Fermi surfaces, we restrict ourselves to work within a thin shell $\pm\Lambda$ around $E_F$ similarly as previous works~\cite{roldanInteractionsSuperconductivityHeavily2013,shafferCrystallineNodalTopological2020,wanObservationSuperconductingCollective2022,horholdTwobandsIsingSuperconductivity2023,shafferWeakcouplingTheoryPair2023,aaseConstrainedWeakcouplingSuperconductivity2023,royUnconventionalPairingIsing2024}.
The relevant available momentum transfers $\Qm(q)$ are shown in \cref{fig: matrix elements FStoFS}.
Intra-pocket scattering for the $\Gamma$ pocket ($\Gamma\Gamma$) contains both low momentum and high momentum contributions.
The dominant contribution to intra-pocket scattering for $K/K'$-type pockets ($KK$) are from low momentum exchange as compared to higher momentum exchanges for inter-pocket ($KK'$) scattering.
The inter-pocket scattering from $\Gamma$ to $K/K'$ (denoted $\Gamma K$) is highly anisotropic and dominated by higher momentum transfers.

\section{Superconductivity in monolayer \texorpdfstring{NbSe$_2$}{NbSe2} \label{model for sc in nbse2}}
As outlined in the main text, we consider only pairing between time-reversal symmetry (TRS) partners ($\bm{k},\sigma,-\bm{k}=:\bar{\bm{k}},-\sigma=:\bar{\sigma}$).
Finite momentum pairing~\cite{larkinNonuniformStateSuperconductors1964,fuldeSuperconductivityStrongSpinExchange1964,shafferWeakcouplingTheoryPair2023} is in general possible, but not expected for free-standing ML NbSe$_2$~\cite{shafferCrystallineNodalTopological2020}.  
The pairing between these states is captured by the BCS-like Hamiltonian~\cite{bardeenTheorySuperconductivity1957}
\begin{align}
    \label{singleband interacting Hamiltonian trs}
    \hat{H}=\sum_{\bk,\sigma} \xi_{\bk,\sigma}\opc_{\bk,\sigma}^\dagger \opc_{\bk,\sigma}+\frac{1}{2}\sum_{\bk,\bk',\sigma} V_{\bk,\bk',\sigma} \opc_{\bk,\sigma}^\dagger \opc_{\bar{\bk},\bar{\sigma}}^\dagger \opc_{\bar{\bk}',\bar{\sigma}} \opc_{\bk',\sigma}\,.
   \end{align}
Using the mean field approximation on this Hamiltonian, the relevant order parameter associated with the band of spin $\sigma$ is given by the self-consistent expression
\begin{align}
    \label{pairing matrix}
    \Delta_{\bk,\sigma}=-\sum_{\bm{k}'}V_{\bk,\bk',\sigma}\langle \opc_{\bar{\bk}',\bar{\sigma}} \opc_{\bk',\sigma}\rangle\,,
\end{align}
with the corresponding mean field Hamiltonian
\begin{align}
    \hat{H}^{\T{MF}}&=\sum_{\bk,\sigma} \xi_{\bk,\sigma} \opc_{\bk,\sigma}^\dagger \opc_{\bk,\sigma} -\frac{1}{2} \sum_{\bk,\sigma}[ \opc_{\bk,\sigma}^\dagger \opc_{\bar{\bk},\bar{\sigma}}^\dagger \Delta_{\bk,\sigma} + \T{h.c.}]\nl
    &-\frac{1}{2}\sum_{\bk,\bk',\sigma}\langle \opc_{\bk,\sigma}^\dagger \opc_{\bar{\bk},\bar{\sigma}}^\dagger \rangle V_{\bk,\bk',\sigma,} \langle \opc_{\bar{\bk}',\bar{\sigma}} \opc_{\bk',\sigma}\rangle\,.
    \label{hamiltonian mean field}
\end{align}
This Hamiltonian contains terms breaking particle conservation which can conveniently be accounted for by going to Nambu space with basis $\opPsi_{\bk}^{\dagger}=( \opc_{\bk,\up}^{\dagger}, \opc_{\bk,\down}^\dagger , \opc_{\bar{\bk},\up},\opc_{\bar{\bk},\down})$.
The resulting Hamiltonian is given by
\begin{align}
    \label{hamiltonian nambu}
    \hat{H}^{\T{MF}}&=\frac{1}{2}\sum_{\bk} \opPsi_{\bk}^\dagger H^{\T{N}}_{\bk} \opPsi_{\bk} + \frac{1}{2}\sum_{\bk,\sigma}\xi_{\bk,\sigma}\\
    &-\frac{1}{2}\sum_{\bk,\bk',\sigma}\langle \opc_{\bk,\sigma}^\dagger \opc_{\bar{\bk},\bar{\sigma}}^\dagger \rangle V_{\bk,\bk',\sigma} \langle \opc_{\bar{\bk}',\bar{\sigma}} \opc_{\bk',\sigma}\rangle\,,\nonumber
\end{align}
and the Bogoliubov-de Gennes Hamiltonian matrix takes the form
\begin{align}
    \label{hamitlonian matrix nambu}
    H^{\T{N}}_{\bk}= &
    \begin{pmatrix}
    \begin{matrix}
        \xi_{\bk,\up} & \\
        & \xi_{\bk,\down}
    \end{matrix}
    &- \Delta(\bk)\\
    -\Delta^\dagger(\bk)& 
    \begin{matrix}
        -\xi_{\bar{\bk},\up} & \\
        & -\xi_{\bar{\bk},\down}
    \end{matrix}\\
\end{pmatrix}.
\end{align}
The pairing matrix $\Delta(\bk)$ can, in the presence of time-reversal symmetry (TRS), be split as 
\begin{align}
    \label{split singlet triplet}
   \Delta(\bk)&=-\Delta^\top(\bar{\bk})=(\Delta_{s}(\bk)\sigma_0 + \Delta_z(\bk)\sigma_z)i\sigma_y\,,
\end{align}
with $\Delta_s$ and $\Delta_z$ the spin singlet and triplet $z$ component, respectively, with $\sigma_i$ the Pauli matrices.
In the following we use the notation $(\Delta(\bk))_{\sigma,\bar{\sigma}}=:\Delta_{\bk,\sigma}$ for the symmetry-allowed components of the pairing matrix.
Application of magnetic field or spontaneous breaking of TRS as discussed by \citeauthor{shafferCrystallineNodalTopological2020} for strong Rashba SOC would enable a pairing also in the triplet $x$ and $y$ components, e.g., by singlet-triplet conversion~\cite{mockliMagneticfieldInduced2019,shafferCrystallineNodalTopological2020}.
The Hamiltonian in \cref{hamiltonian nambu} can be diagonalized by the Bogoliubov transformation
\begin{align}
    \label{simplified bogoliubov transformation}
    U^{\T{N}}_{\bk}&=
    \begin{pmatrix}
        u_{\bk,\up} & 0 & 0 & v_{\bk,\up}\\
        0 & u_{\bk,\down} & v_{\bk,\down} & 0 \\
        0 & -v_{\bk,\down} & u_{\bk,\down} & 0 \\
        -v_{\bk,\up} & 0 & 0 & u_{\bk,\up}\\
    \end{pmatrix}   
    ,\\
    U^{\T{N}}_{\bk}H^{\T{N}}_{\bk}(U^{\T{N}}_{\bk})^{-1}&=\T{diag}(E_{\bk,\up},E_{\bk,\down},-E_{\bar{\bk},\up},-E_{\bar{\bk},\down}),
\end{align}
with ${E_{\bk,\sigma}=\sqrt{\xi_{\bk,\sigma}^2+|\Delta_{\bk,\sigma}^2|}}$ and the usual Bogoliubov transformation parameters~\cite{bogoljubovNewMethodTheory1958}
\begin{align}
    \label{boboliubov transformation}
    u_{\bk,\sigma}=\sqrt{\frac{1}{2}\left(1+\frac{\xi_{\bk,\sigma}}{E_{\bk,\sigma}}\right)},\\
    v_{\bk,\sigma}=\frac{-\Delta_{\bk,\sigma}}{|\Delta_{\bk,\sigma}|}\sqrt{\frac{1}{2}\left(1-\frac{\xi_{\bk,\sigma}}{E_{\bk,\sigma}}\right)}\,.
\end{align}
Inserting this transformation yields the Bogoliubov quasiparticle operators  
\begin{align}
\label{generalized bogoliubov transformation}
    &(\opg_{\bk,\up}, \opg_{\bk,\down}, \opg_{\bar{\bk},\up}^\dagger, \opg_{\bar{\bk},\down}^\dagger,)^{\top}=U^{\T{N}}_{\bk} \opPsi\,,\\
    &\{\opg_{\bk,\sigma},\opg_{\bk',\sigma'}^\dagger\}=\delta_{\bk,\bk'}\delta_{\sigma,\sigma'}\,, \\
    &\{\opg_{\bk,\sigma},\opg_{\bk',\sigma'}\}=0\,,
\end{align}
in terms of which \cref{hamiltonian nambu} can be written as
\begin{align}
    \label{hamiltonian diagonal nambu}
    \hat{H}^{\T{MF}}&=
    \frac{1}{2}\sum_{\bk,\sigma} E_{\bk,\sigma}\opg_{\bk,\sigma}^\dagger \opg_{\bk,\sigma} - \frac{1}{2}\sum_{\bk,\sigma}(E_{\bk,\sigma}-\xi_{\bk,\sigma})\nl
    &-\frac{1}{2}\sum_{\bk,\bk',\sigma}\langle \opc_{\bk,\sigma}^\dagger \opc_{\bar{\bk},\bar{\sigma}}^\dagger \rangle V_{\bk,\bk',\sigma}\langle \opc_{\bar{\bk}',\bar{\sigma}} \opc_{\bk',\sigma}\rangle\,.
\end{align}
This diagonalization fixes the thermal expectation value occurring in the definition of the pairing matrix in \cref{pairing matrix} to
\begin{align}
    \label{expectation value kramers pair thermal}
    \langle \opc_{\bar{\bk},\bar{\sigma}} \opc_{\bk,\sigma} \rangle=
    \frac{\tanh(\beta E_{\bk,\sigma}/2)}{2E_{\bk,\sigma}}\Delta_{\bk,\sigma}=:\Pi(E_{\bk,\sigma})\Delta_{\bk,\sigma}\,.
\end{align}
Inserting \cref{expectation value kramers pair thermal} into \cref{pairing matrix} yields the familiar BCS gap equation \cref{main-gap equation trs}.

Using \cref{pairing matrix} and \cref{expectation value kramers pair thermal} in \cref{hamiltonian diagonal nambu}, we obtain the free energy of the superconductor as in \cref{main-free energy}. We emphasize that this gap can only be used to rank the relative stability of the physical solutions, not to find them by optimization with respect to $\Delta$~\cite{hutchinsonMixedTemperaturedependentOrder2021,aaseConstrainedWeakcouplingSuperconductivity2023}.

\subsection{Temperature dependence of the gaps for selected basis functions\label{subsec: expansion coefficients}}

To use the projection method discussed in the main paper, we first set up the basis functions required.
The symmetry group of freestanding ML NbSe$_2$ is $D_{3\T{h}}$.
However, due to the symmetry of orbitals making up the Nb derived bands under $\sigma_{\T{h}}$, the irreducible representations $A_1^{\prime\prime}$, $A_2^{\prime}$ and $E^{\prime\prime}$, which are antisymmetric under this operation, do not contribute to the expansion \cref{main-expansion gap} with respect to the-in plane momentum $\bk$. 
Removing these irreducible representations from the expansion, we are left with an effective symmetry group $D_3$ describing the in-plane behavior of the gap.
As such, we drop the primes for the remaining irreducible representations when convenient for notation.

To construct a set of basis functions for each irreducible representation we employ the symmetrization procedure from \citeauthor{hanisDistinguishingNodalNonunitary2024}~\cite{hanisDistinguishingNodalNonunitary2024}.
There, the authors make use of a projection formalism to extract from a seed function with the appropriate periodicity of the lattice the component that transforms according to the desired irreducible representation.
After generating candidate functions with the above method, we employ a Gram-Schmidt procedure to ensure the orthonormality property \cref{main-orthogonaity basis functions} on each Fermi surface.
Due to the different transformation behavior under $D_3$, the basis functions of each irreducible representation only need to be orthonormalized with respect to functions in the same irreducible representation.
We find from the linearized gap equation that pairing with symmetries from the $A'_1$ and $A_2''$ irreducible representations has negligible strength when compared to the $E'$ irreducible representation.
Furthermore, inclusion of basis functions of the  $A'_1$ and $A''_2$ irreducible representation resulted in negligible contributions to the two leading solutions discussed in the main text.
As such we now focus solely on the $E'$ irreducible representation for the following expansion.

\nolinenumbers
\begin{figure*}[ht]
    \centering
    \vspace{-4ex}
    \subfloat[]
    {
        \include{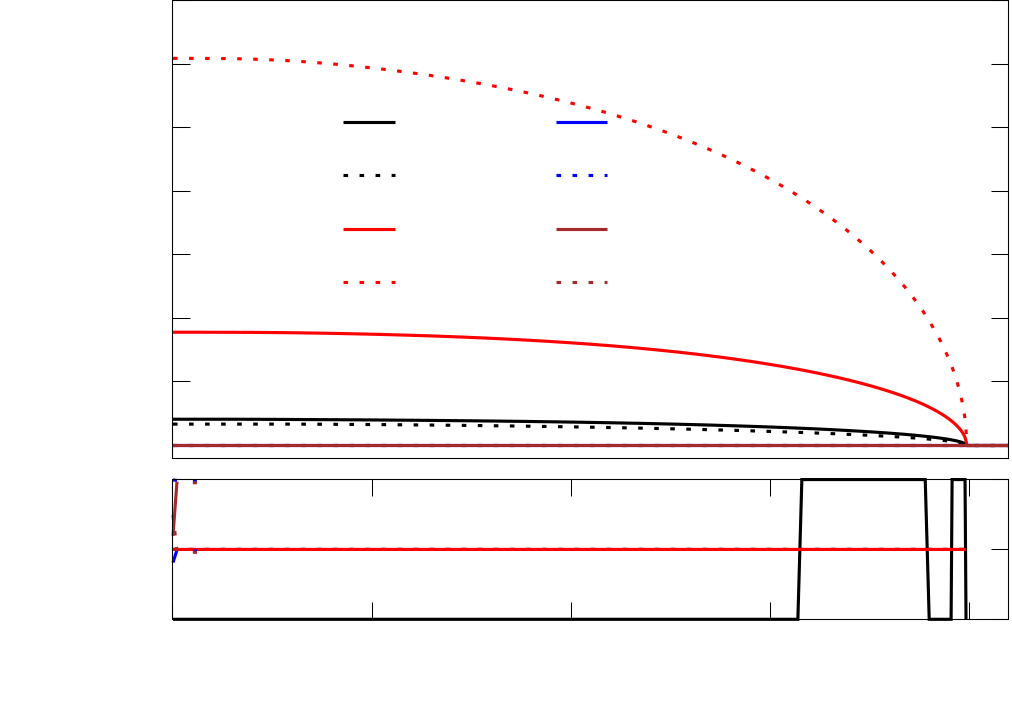}
    }
    \subfloat[]
    {
        \include{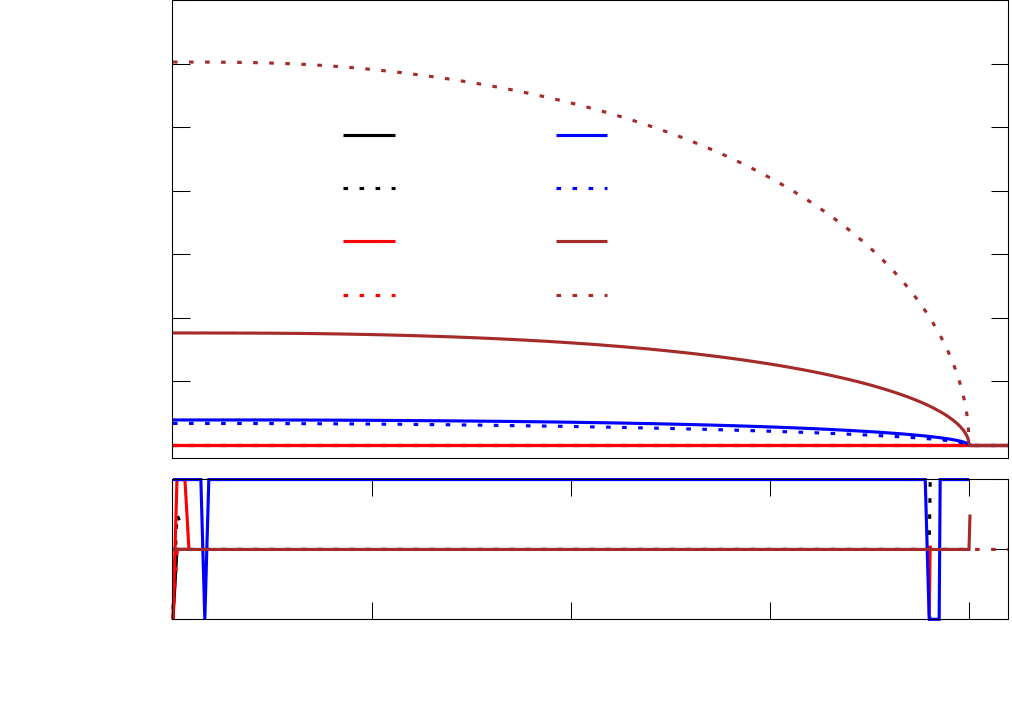}
    }\\
    \vspace{-2ex}
    \subfloat[]
    {
        \include{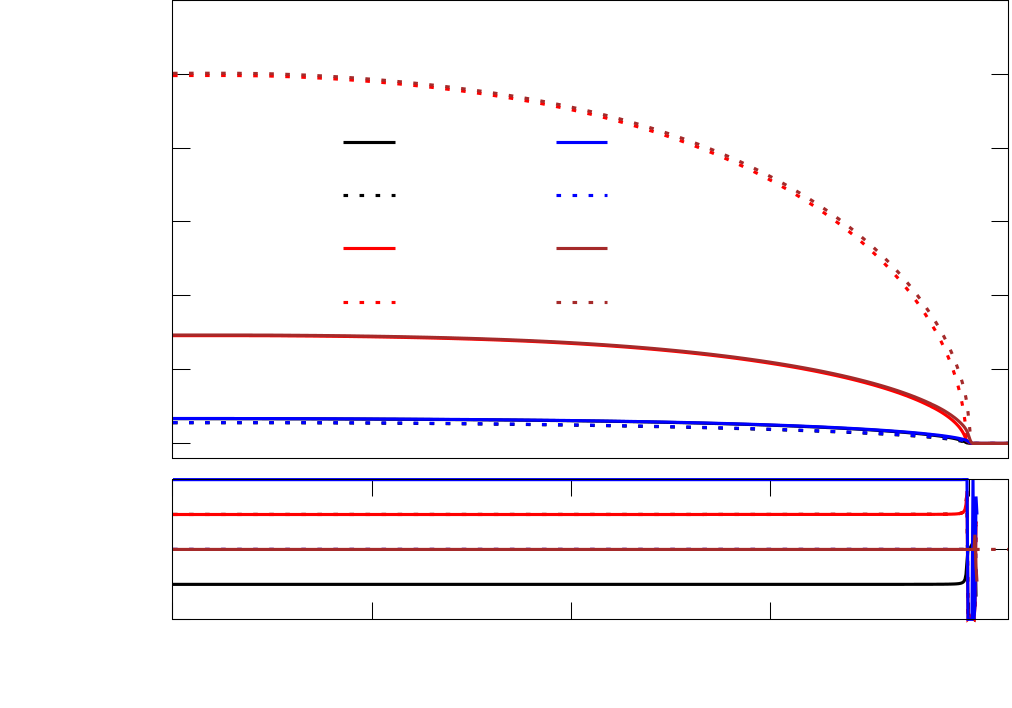}
    }
    \vspace{-2ex}
    \caption{
    Temperature dependence of the expansion coefficients for different superconducting phases as obtained from the self consistent gap equation.
    The nematic $p_x$ and $p_y$ solutions shown in \cref{main-fig: linearized gaps} as obtained using the linearized equations are evolved down to $T=0\,\T{K}$ with the respective modulus and phase of their expansion coefficients shown in panels (a) and (b) respectively.
    The complex superposition of these nematic solutions yields a chiral solution which is shown in panel (c).
    The gauges of the different solutions are chosen such as to keep the expansion component with largest modulus purely real.
    }
    \label{fig: T dependence}
\end{figure*}

We start with the seed function $f(\bk)=\exp(i\bk\cdot\bm{a}_1)$ and project out the respective spin singlet and $z$-triplet components transforming as $E'$.
Since the $E'$ irreducible representation is two-dimensional, a simple way to generate sufficiently many unique basis functions is given by taking the real and imaginary parts of the resulting complex functions generated by the above method.
The resulting basis functions are
\begin{align}
    f^\pi_{Es1}&=\frac{1}{\mathcal{N}^{\pi}_{Es1}}\left(\cos(k_x a) - \cos(k_x a/2)\cos(k_y a\sqrt{3}/2)\right)\,,\\
    f^\pi_{Es2}&=\frac{1}{\mathcal{N}^{\pi}_{Es2}}\left(\sin(k_x a/2) \sin(k_y a\sqrt{3}/2)\right)\,,
\end{align}
for the singlets where we normalize them using
\begin{align}
    \mathcal{N}^{\pi}_{\mu}&=\sqrt{\langle f^{\pi}_{\mu},f^{\pi}_{\mu}\rangle_\pi}\,.
\end{align} 
The $z$-triplet basis functions we find by the above method are 
\begin{align}
    f_{Ez1}&=\left(\sin(k_x a) + \sin(k_x a/2)\cos(k_y a\sqrt{3}/2)\right)\,,\\
    f_{Ez2}&=\cos(k_x a/2) \sin(k_y a\sqrt{3}/2)\,.
\end{align}
Note that the symmetry under $C_2$ with respect to the $x$-axis ensures that the real and imaginary parts of the singlet and triplet components are already orthogonal to each other, however due to the presence of SOC the singlet and $z$-triplet components are mixed.
To restore orthonormality, we use the Gram-Schmidt procedure by subtracting the scalar product between singlet and $z$-triplet components as
\begin{align}
    f^\pi_{Ez1}&=\frac{1}{\mathcal{N}^{\pi}_{Ez1}}\left(f_{Ez1}-\frac{\langle f_{Ez1},f^\pi_{Es1}\rangle_\pi}{\langle f^\pi_{Es1},f^{\pi}_{Es1}\rangle_\pi}f^{\pi}_{Es1}\right)\,,\\
    f^\pi_{Ez2}&=\frac{1}{\mathcal{N}^{\pi}_{Ez2}}\left(f_{Ez2}-\frac{\langle f_{Ez2},f^\pi_{Es2}\rangle_\pi}{\langle f^\pi_{Es2},f^{\pi}_{Es2}\rangle_\pi}f^{\pi}_{Es2}\right)\,.
\end{align}

We show the temperature dependence of the nematic phases $p_y$ and $p_x$ in \cref{fig: T dependence}(a \& b).
The nematicity of these phases results in three equivalent solutions for $p_y$ and $p_x$, which differ by a rotation of their nodal line by $C_3$ symmetry.
The temperature dependence of the chiral solution, obtained by allowing for complex coefficients $\Delta_\mu^\pi$, is shown in \cref{fig: T dependence}(f).
Note also that chiral phases corresponding to either $p_x \pm i p_y$ are equivalent solutions of the gap equations with the system chosing either by spontaenous time-reversal symmetry breaking.

\subsection{Chemical potential and external screening\label{sec: dependence on chemical potential}}

While our theoretical calculations considered free-standing monolayer NbSe$_2$, experiments are usually performed on mono- or few-layer NbSe$_2$ samples deposited or grown on a substrate~\cite{xiIsingPairingSuperconducting2016,hamillTwofoldSymmetricSuperconductivity2021,kuzmanovicTunnelingSpectroscopyFewmonolayer2022}.
In addition, such samples are often encapsulated from the top as well in order to protect against oxidation. 
Thus, in most experiments NbSe$_2$ is inherently subject to the impact of an environment that modifies the screening discussed in the main text.
Different substrates have been shown to \hbox{modify} ordered phase realized in few-layer NbSe$_2$, by selectively suppressing either the emergence of an incommensurate charge density wave phase, superconductivity or both~\cite{dreherProximityEffectsCharge2021}.
As we are focusing on pairing due to the screened Coulomb interaction in this material, we discuss here briefly the impact of gating and screening by the environment surrounding the ML.

\begin{figure}[ht]
\centering
\include{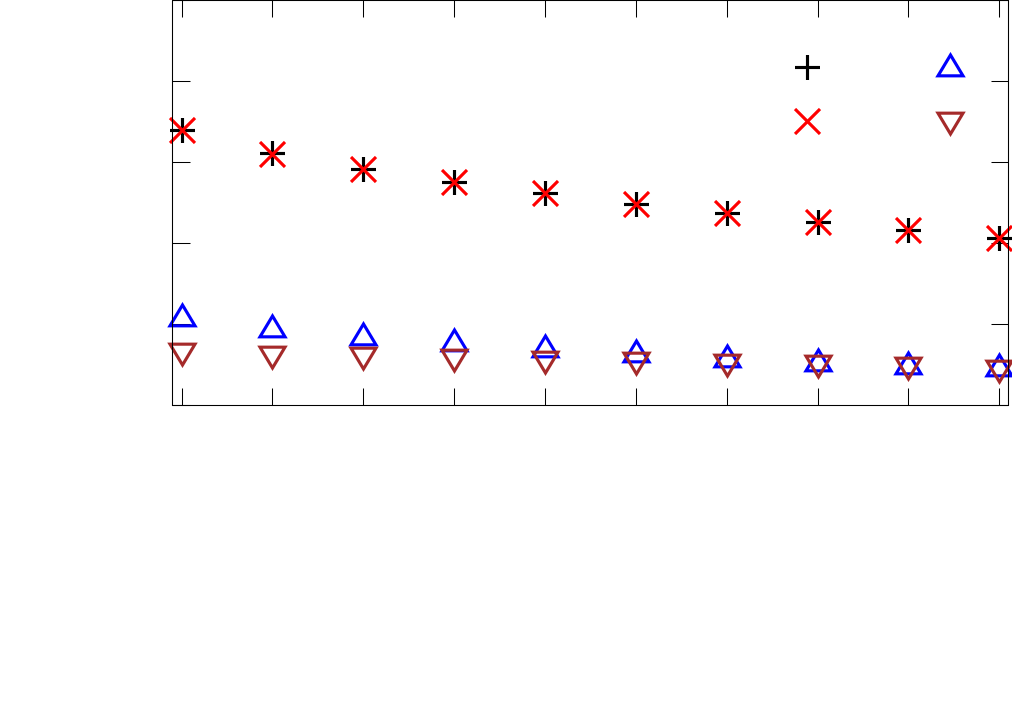}
\caption{\label{fig: Tc on epsr} Decrease of the eigenvalues of the linearized gap equation with increasing dielectric constant of the environment.}
\end{figure}

The static screening of a dielectric environment can be taken into account at the level of \cref{main-formal solution dyson equation in 2d,main-dielectric tensor} by introducing a nontrivial dielectric constant $\epsilon_r$ into $V_0^{\T{2D}}$.
Increased dielectric screening by the substrate suppresses the Coulomb interactions overall, and it leads to monotonically decreasing $T_{\T{c}}$ for all gap symmetries, as shown in Fig.~\ref{fig: Tc on epsr}.
\nolinenumbers
The situation is more complicated for semi-/metallic environments, where an additional density of free charge carriers is available to contribute to the screening at low-momentum exchanges.
However, the impact of this type of environment at larger momentum transfers is strongly dependent on the microscopic details of the specific setup under consideration and as such beyond the scope of this work.

\begin{figure}[ht]
    \centering
        \includegraphics[width=\columnwidth]{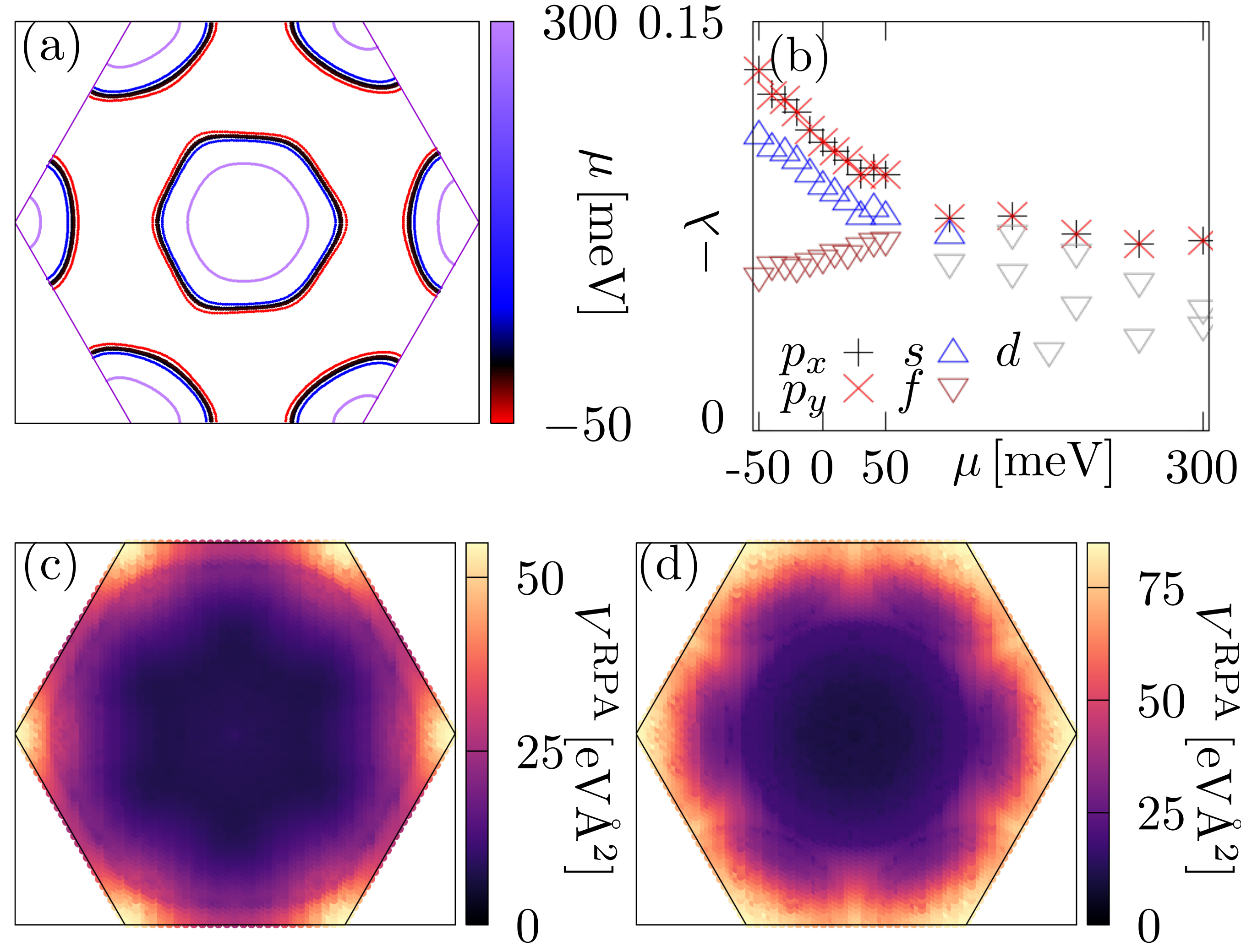}
    \caption{\label{fig: variation with mu}
    Influence of varying chemical potential on superconducting order. Shown are the changing Fermi surfaces (a), pairing strengths for the leading superconducting instabilities (b) and interaction strengths at chemical potential $-50\,\T{meV}$ (c) and $300\,\T{meV}$ (d), respectively. 
    }    
\end{figure}

Gating the monolayer can either be pursued actively, e.g., by liquid ion gating~\cite{xiGateTuningElectronic2016}, or be a passive effect due to a charge transfer between the substrate and the ML caused by a difference in their work functions.
In the following we have concentrated on the values of $\mu=\pm50$~meV motivated by \cite{xiGateTuningElectronic2016} and $\mu=300$~meV seen in a misfit layer compound of LaSe/NbSe$_2$~\cite{lericheMisfitLayerCompounds2021}.
Regardless of the origin of the gating, it affects the superconducting pairing chiefly in two distinct ways.
Firstly, changing the chemical potential will modify the shape and orbital composition of the Fermi surfaces where the \hbox{dominant} contribution to the pairing occurs.
This change is shown in \cref{fig: variation with mu}(a).
Simultaneously, the change in the Fermi surfaces will also affect the \hbox{polarizability} and with it the screened interaction potential.
Together, the change in the pairing strength can be non-monotonous as illustrated by the behavior of the leading nematic solutions $p_x$ and $p_y$.
The complex interplay between weaker screening and shrinking Fermi surfaces at higher chemical potentials is highlighted by comparing the screened interactions at $\mu=-50\,\T{meV}$, $\mu=300\,\T{meV}$ and $\mu=0\,\T{meV}$ from the main text.
The screened interaction for the former two is shown in \cref{fig: variation with mu}(c) and (d), respectively.
As expected the higher density of states for $\mu=-50\,\T{meV}$ results in increased screening and lower interaction strengths throughout the BZ.
Conversely, the screened interaction is stronger at $\mu=300\,\T{meV}$.

\clearpage

\makeatletter\@input{main2supp.tex}\makeatother

\bibliography{references}